\documentclass[tighten,twocolumn]{aastex701}
\usepackage{csvsimple}
\usepackage{multirow}
\usepackage{rotating}
\usepackage{amsmath}
\usepackage{longtable}
\usepackage{booktabs}
\usepackage{ltablex}
\usepackage{epsfig}
\usepackage{threeparttable}
\usepackage{url}

\def\nustar{\emph{NuSTAR}}

\def\ep{\emph{Einstein Probe}}

\def\flux{\mbox{erg\,cm$^{-2}$\,s$^{-1}$}}
\def\lum{\mbox{erg\,s$^{-1}$}}

\def\cm2{\rm \ cm$^{-2}$}

\def\arcsec{\mbox{$^{\prime\prime}$}}
\def\arcmin{\mbox{$^{\prime}$}}
\def\deg{\mbox{$^{\circ}$}}

\def\msun{~M_{\odot}}

\newcommand\T{\rule{0pt}{2.6ex}}       
\newcommand\B{\rule[-1.2ex]{0pt}{0pt}}

\def\srcfirst{EP\,J171159.4$-$333253}
\def\src{EP\,J1711$-$3332}

\shorttitle{EP\,J171159.4$-$333253: a rare clocked burster with eclipses}
\shortauthors{Wang et al.}

\begin{document}

\title{Einstein Probe discovery of EP\,J171159.4$-$333253: \\an eclipsing neutron star low-mass X-ray binary with clocked bursts}

\correspondingauthor{Yilong Wang}

\author[0009-0009-1721-3663]{Y.~L.~Wang}
\affiliation{National Astronomical Observatories, Chinese Academy of Sciences, 20A Datun Road, Beijing 100101, China}
\affiliation{Institute of Space Sciences (ICE, CSIC), Campus UAB, Carrer de Can Magrans s/n, Barcelona, E-08193, Spain}
\affiliation{Institut d’Estudis Espacials de Catalunya (IEEC), Carrer Gran Capità 2–4, Barcelona, E-08034, Spain}
\affiliation{School of Astronomy and Space Science, University of Chinese Academy of Sciences, 19A Yuquan Road, Beijing 100049, China}
\email[show]{wangyilong@nao.cas.cn}

\author[0000-0001-7611-1581]{F.~Coti~Zelati}
\affiliation{Institute of Space Sciences (ICE, CSIC), Campus UAB, Carrer de Can Magrans s/n, Barcelona, E-08193, Spain}
\affiliation{Institut d’Estudis Espacials de Catalunya (IEEC), Carrer Gran Capità 2–4, Barcelona, E-08034, Spain}
\affiliation{INAF--Osservatorio Astronomico di Brera, Via Bianchi 46, I-23807 Merate (LC), Italy}
\email[]{cotizelati@ice.csic.es}  

\author[0000-0002-0430-6504]{E.~Parent}
\affiliation{Institute of Space Sciences (ICE, CSIC), Campus UAB, Carrer de Can Magrans s/n, Barcelona, E-08193, Spain}
\affiliation{Institut d’Estudis Espacials de Catalunya (IEEC), Carrer Gran Capità 2–4, Barcelona, E-08034, Spain}
\email[]{parent@ice.csic.es}

\author[0000-0001-5674-4664]{A.~Marino}
\affiliation{Institute of Space Sciences (ICE, CSIC), Campus UAB, Carrer de Can Magrans s/n, Barcelona, E-08193, Spain}
\affiliation{Institut d’Estudis Espacials de Catalunya (IEEC), Carrer Gran Capità 2–4, Barcelona, E-08034, Spain}
\affiliation{INAF, Istituto di Astrofisica Spaziale e Fisica Cosmica, Via U. La Malfa 153, I-90146 Palermo, Italy}
\email[]{marino@ice.csic.es}

\author[0000-0003-2177-6388]{N.~Rea}
\affiliation{Institute of Space Sciences (ICE, CSIC), Campus UAB, Carrer de Can Magrans s/n, Barcelona, E-08193, Spain}
\affiliation{Institut d’Estudis Espacials de Catalunya (IEEC), Carrer Gran Capità 2–4, Barcelona, E-08034, Spain}
\email[]{rea@ice.csic.es}

\author[0000-0003-4236-9642]{V.~S.~Dhillon}
\affiliation{Astrophysics Research Cluster, School of Mathematical \& Physical Sciences, University of Sheffield, Sheffield S3 7RH, UK}
\affiliation{Instituto de Astrof\'{\i}sica de Canarias, E-38205 La Laguna, Tenerife, Spain}
\email[]{vik.dhillon@sheffield.ac.uk}

\author[]{J.~Blanco-Pozo}
\affiliation{Institute of Space Sciences (ICE, CSIC), Campus UAB, Carrer de Can Magrans s/n, Barcelona, E-08193, Spain}
\affiliation{Institut d’Estudis Espacials de Catalunya (IEEC), Carrer Gran Capità 2–4, Barcelona, E-08034, Spain}
\email[]{blanco@ice.csic.es}

\author[0000-0002-6689-0312]{I.~Ribas}
\affiliation{Institute of Space Sciences (ICE, CSIC), Campus UAB, Carrer de Can Magrans s/n, Barcelona, E-08193, Spain}
\affiliation{Institut d’Estudis Espacials de Catalunya (IEEC), Carrer Gran Capità 2–4, Barcelona, E-08034, Spain}
\email[]{iribas@ice.csic.es}

\author[0000-0001-7221-855X]{S.~P.~Littlefair}
\affiliation{Astrophysics Research Cluster, School of Mathematical \& Physical Sciences, University of Sheffield, Sheffield S3 7RH, UK}
\email[]{s.littlefair@sheffield.ac.uk}

\author[0009-0006-0436-2975]{Z.~H.~Yang}
\affiliation{Key Laboratory of Particle Astrophysics, Institute of High Energy Physics, Chinese Academy of Sciences, Beijing 100049, China}
\affiliation{School of Astronomy and Space Science, University of Chinese Academy of Sciences, 19A Yuquan Road, Beijing 100049, China}
\email[]{yangzh@ihep.ac.cn}

\author[0000-0001-8630-5435]{G.~B.~Zhang}
\affiliation{Yunnan Observatories, Chinese Academy of Sciences, Kunming 650216, China}
\affiliation{Key Laboratory for the Structure and Evolution of Celestial Objects, Chinese Academy of Sciences, Kunming 650216, China}
\email[]{zhangguobao@ynao.ac.cn}

\author[0000-0002-6449-106X]{S.~Guillot}
\affiliation{Institut de Recherche en Astrophysique et Plan\'{e}tologie, UPS-OMP, CNRS, CNES, 9 avenue du Colonel Roche, BP 44346, Toulouse Cedex 4, 31028, France}
\email[]{sebastien.guillot@irap.omp.eu}

\author[0009-0000-0015-9124]{K.~R.~Ni}
\affiliation{Institute of Astrophysics, Central China Normal University, Wuhan 430079, China}
\email[]{nikangrui0@163.com}

\author[0009-0003-1518-6186]{J.~H.~Wu}
\affiliation{Department of Astronomy, Guangzhou University, Guangzhou 510006, China}
\email[]{jhwu@e.gzhu.edu.cn}

\author[0000-0002-6459-0674]{A.~Patruno}
\affiliation{Institute of Space Sciences (ICE, CSIC), Campus UAB, Carrer de Can Magrans s/n, Barcelona, E-08193, Spain}
\affiliation{Institut d’Estudis Espacials de Catalunya (IEEC), Carrer Gran Capità 2–4, Barcelona, E-08034, Spain}
\email[]{patruno@ice.csic.es}

\author[0000-0002-6447-3603]{Y.~Cavecchi}
\affiliation{Departamento de Astrof\'isica, Universidad de La Laguna, 38206, San Crist\'obal de La Laguna, Tenerife, Spain}
\affiliation{Instituto de Astrof\'{\i}sica de Canarias, E-38205 La Laguna, Tenerife, Spain}
\affiliation{Departament de Fis\'{i}ca, EEBE, Universitat Polit\`ecnica de Catalunya, Av. Eduard Maristany 16, 08019 Barcelona, Spain}
\affiliation{Center for Nuclear Astrophysics across Messengers (CeNAM), 640 S Shaw Lane, East Lansing, MI 48824, USA}
\email[]{yuri.cavecchi@upc.edu}

\author[0000-0003-4795-7072]{G.~Illiano}
\affiliation{INAF--Osservatorio Astronomico di Brera, Via Bianchi 46, I-23807 Merate (LC), Italy}
\email[]{giulia.illiano@inaf.it}

\author[0000-0001-6289-7413]{A.~Papitto}
\affiliation{INAF--Osservatorio Astronomico di Roma, Via Frascati 33, I-00078 Monte Porzio Catone (RM), Italy}
\email[]{alessandro.papitto@inaf.it}

\author[0000-0001-7915-996X]{F.~Ambrosino}
\affiliation{INAF--Osservatorio Astronomico di Roma, Via Frascati 33, I-00078 Monte Porzio Catone (RM), Italy}
\email[]{filippo.ambrosino@inaf.it}

\author[0000-0001-9920-4019]{B.~F.~Liu}
\affiliation{National Astronomical Observatories, Chinese Academy of Sciences, 20A Datun Road, Beijing 100101, China}
\affiliation{School of Astronomy and Space Science, University of Chinese Academy of Sciences, 19A Yuquan Road, Beijing 100049, China}
\email[]{bfliu@nao.cas.cn}

\author[0000-0003-4200-9954]{H.~Q.~Cheng}
\affiliation{National Astronomical Observatories, Chinese Academy of Sciences, 20A Datun Road, Beijing 100101, China}
\email[]{hqcheng@bao.ac.cn}

\author[0000-0001-7584-6236]{H.~Feng}
\affiliation{Key Laboratory of Particle Astrophysics, Institute of High Energy Physics, Chinese Academy of Sciences, Beijing 100049, China}
\email[]{hfeng@ihep.ac.cn}

\author[0000-0002-0779-1947]{J.~W.~Hu}
\affiliation{National Astronomical Observatories, Chinese Academy of Sciences, 20A Datun Road, Beijing 100101, China}
\email[]{hujingwei@nao.cas.cn}

\author[0000-0002-2006-1615]{C.~C.~Jin}
\affiliation{National Astronomical Observatories, Chinese Academy of Sciences, 20A Datun Road, Beijing 100101, China}
\affiliation{School of Astronomy and Space Science, University of Chinese Academy of Sciences, 19A Yuquan Road, Beijing 100049, China}
\affiliation{Institute for Frontier in Astronomy and Astrophysics, Beijing Normal University, Beijing 102206, China}
\email[]{ccjin@nao.cas.cn}

\author[0000-0002-9615-1481]{H.~Sun}
\affiliation{National Astronomical Observatories, Chinese Academy of Sciences, 20A Datun Road, Beijing 100101, China}
\email[]{hsun@bao.ac.cn}

\author[0000-0002-2705-4338]{L.~Tao}
\affiliation{Key Laboratory of Particle Astrophysics, Institute of High Energy Physics, Chinese Academy of Sciences, Beijing 100049, China}
\email[]{taolian@ihep.ac.cn}

\author[]{Y.~J.~Xu}
\affiliation{Key Laboratory of Particle Astrophysics, Institute of High Energy Physics, Chinese Academy of Sciences, Beijing 100049, China}
\email[]{xuyj@ihep.ac.cn}

\author[0000-0002-7680-2056]{H.~N.~Yang}
\affiliation{National Astronomical Observatories, Chinese Academy of Sciences, 20A Datun Road, Beijing 100101, China}
\email[]{hnyang@nao.cas.cn}

\author[]{W.~Yuan}
\affiliation{National Astronomical Observatories, Chinese Academy of Sciences, 20A Datun Road, Beijing 100101, China}
\affiliation{School of Astronomy and Space Science, University of Chinese Academy of Sciences, 19A Yuquan Road, Beijing 100049, China}
\email[]{wmy@nao.cas.cn}

\author[0000-0001-9893-8248]{Q.~C.~Zhao}
\affiliation{Key Laboratory of Particle Astrophysics, Institute of High Energy Physics, Chinese Academy of Sciences, Beijing 100049, China}
\affiliation{School of Astronomy and Space Science, University of Chinese Academy of Sciences, 19A Yuquan Road, Beijing 100049, China}
\email[]{zhaoqc@ihep.ac.cn}


\submitjournal{ApJ}

\begin{abstract}
\srcfirst\ is a new neutron-star low-mass X-ray binary discovered in outburst by the \ep\ (EP) on 2025 June 23, exhibiting clocked type-I X-ray bursts, eclipses and dips. In this paper, we report on the results of the X-ray spectral and timing analyses for \srcfirst\ using data collected by EP and \nustar\ during the first 21 days of the outburst. The X-ray burst recurrence time can be characterized over a subset of nine bursts spanning 1.6\,days around the \nustar\ observation, and the result is $t_{\rm rec}=8196 \pm 177\,$s with indications of a possible decreasing trend. From the X-ray eclipse events, the binary orbital period and the eclipse duration are estimated to be $P_{\rm orb}=6.48301 \pm 0.00003$\,hr and $D_{\star,X} = 1245.5^{+6.9}_{-6.5}$\,s, respectively. These enable an estimate of the mass and radius of the companion star and the binary inclination, which are $M_2\approx0.6-0.8\,M_\odot$, $R_2\approx0.7-0.8\,R_\odot$ and $i\approx73-75^\circ$, respectively. We also report on joint ULTRACAM and EP observations on 2025 July 21--22, detecting the source optical counterpart and covering an eclipse in both X-ray and optical bands. The optical eclipse is wavelength-dependent and broader than in X-rays, indicating that part of the optical emission arises from an extended region in the accretion flow. Despite a moderate variation in the source flux, the properties of the persistent X-ray emission are typical of a hard spectral state. We further evaluated the ratio of the accretion energy to the thermonuclear energy to be 120--130, implying helium bursts with the accreted hydrogen being depleted in-between bursts. 
\end{abstract}



\section{Introduction}
\label{sec:intro}
Low-mass X-ray binaries (LMXBs) are binary systems composed of a compact stellar object, either a black hole (BH) or a neutron star (NS), and a companion star typically with a mass comparable to or less than the solar mass \citep[e.g.][]{lewin1995,Bahramian2022}. In NS-LMXBs, the matter accreted on top of the NS surface can sometimes undergo thermonuclear runaways and produce explosions, known as type-I X-ray bursts \citep[e.g.][]{Fujimoto1981,Galloway2021}. These phenomena naturally require the accreting compact object to possess a surface where matter can accumulate, making type-I X-ray bursts exclusive to NS-LMXBs and thus a clear signature of an NS accretor. Type-I X-ray bursts are usually observed to repeat at variable intervals, depending on the accretion rate of the system as well as the chemical composition of the accreted matter. Among more than one hundred previously known bursters \citep{Galloway2020}, only seven sources are known to have shown bursts that repeat quasi-periodically, which are referred to as \emph{clocked} bursters: GS\,1826$-$24 \citep[the prototype; e.g.][]{Ubertini1999}, KS\,1731$-$260 \citep{Cornelisse2003}, GS\,0836$-$429 \citep{Aranzana2016}, IGR\,J17480$-$2446 \citep{Motta2011, Chakraborty2011}, MAXI\,J1816$-$195 \citep{Bult2022, WangPJ2024}, 1RXS\,J180408.9$-$342058 \citep{Marino2019, Fiocchi2019} and SRGA\,J144459.2$-$60420 \citep{Ng2024, Molkov2024, Dohi2025, Papitto2025}. 

In this paper, we report on the discovery of a new clocked burster, \srcfirst\ (hereafter \src). \src\ was first detected by the Wide-field X-ray Telescope (WXT) on board the \ep\ (EP) mission \citep{yuan22, Yuan2025} in one observation performed on 2025 June 23, and therefore was initially designated as EP250623a \citep{Ni2025}. However, stacking of WXT data showed that a weak signal from the source location could already be detected at least on June 21. Then, on June 25, a target-of-opportunity (ToO) observation at the WXT position was performed with the Follow-up X-ray Telescope (FXT) on board EP. An uncatalogued X-ray source was detected at ${\rm R.\,A.\,(J2000)} = 17^{\rm h}11^{\rm m}59.4^{\rm s}$, ${\rm Dec.\,(J2000)} = -33^{\circ}32^{\prime}53^{\prime\prime}$, with an uncertainty of $10\arcsec$ in radius (90\% confidence level, statistical and systematic), giving the source a final designation of \srcfirst\ \citep{Wu2025}. The FXT light curve, apart from the persistent emission, showed a short burst lasting $\sim100$\,s and a low-count-rate period (later identified as a dip event), which, together with the low Galactic latitude of \src, led to the tentative identification of the source as an eclipsing NS-LMXB. A \emph{Nuclear Spectroscopic Telescope Array} (\nustar) follow-up observation was performed on 2025 June 27. Seven bursts with quasi-periodic recurrence time of $\sim 8200$\,s were detected \citep{Marino2025}, suggesting that \src\ is a clocked burster. 

A radio follow-up observation performed with \emph{MeerKAT} on 2025 June 30 revealed a counterpart within the FXT position error, with a 1.3-GHz flux density of $108\pm20\,\mu{\rm Jy}$ \citep{Cowie2025}. In the optical band, on 2025 June 28, a ToO observation with the Visible Telescope (VT) on board the \emph{Space Variable Objects Monitor} (SVOM) detected a candidate optical counterpart with magnitudes of $\sim18.5$~mag and $\sim20.0$~mag in the VT-Red and VT-Blue bands, respectively \citep{Guillot2025}. However, the source was not detected in two observations performed respectively with the 0.4-m \emph{Super Light Telescope} (SLT) and the 0.5-m \emph{Robotic Imager For Transients} (RIFT) at Lulin Observatory on 2025 June 27 and July 3, which gave 3-$\sigma$ upper limits of 18.8~mag with the rp filter of SLT and 19.0~mag with the white filter of RIFT \citep{Kong2025}. On 2025 July 21, we confirmed the optical counterpart of \src\ using the ULTRACAM camera mounted on the 3.5-m \emph{New Technology Telescope} (NTT), as reported in this work. 

This paper is structured as follows: in Section~\ref{sec:data reduction}, we describe the X-ray and optical observations and the data reduction procedures. The results of our data analysis are presented in Section~\ref{sec:results}, and then discussed in Section~\ref{sec:discussion}. Finally, we summarize in Section~\ref{sec:summary}.

\section{Observations and data reduction} \label{sec:data reduction}
Our main dataset includes observations obtained over the first 21 days of the outburst with the following X-ray telescopes: WXT and FXT on board EP and \nustar\ (see Table~\ref{tab:log} for details). Photon arrival times recorded by the different instruments were corrected to the Solar System barycenter using the JPL DE430 ephemeris and the source position determined from \emph{MeerKAT} observations: R.\,A. = 17$^\mathrm{h}$11$^\mathrm{m}$59.31$^\mathrm{s}$, Dec. = –33\deg32\arcmin52.5\arcsec\,(J2000.0; \citealt{Cowie2025}).

Additionally, we present the results of an optical observation performed on 2025 July 21-22 by ULTRACAM, and partial results of a simultaneous X-ray observation by EP/FXT. A detailed analysis of this X-ray observation, together with the remaining $\sim30$ observations performed after 2025 July 13, will be presented in a companion paper (Yang et al., in prep.), which focuses on the long-term evolution of \src\ throughout this outburst.

\subsection{\ep} \label{subsec:ep}

\subsubsection{Wide-field X-ray Telescope (WXT)} \label{subsubsec:WXT}
The WXT on board EP \citep{yuan22} is a wide-field monitor equipped with the novel lobster-eye micro-pore optics, giving it a field of view (FOV) of $\sim3600\,{\rm deg}^2$. Operating in the 0.5--4\,keV band, WXT has a sensitivity of $(2-3)\times10^{-11}\,\flux$ for a typical exposure of 1\,ks. The energy resolution is 170\,eV at 1\,keV and the spatial resolution is $\approx5\arcmin$ (FWHM). 

\begin{figure}
\centering
\includegraphics[width=0.48\textwidth]{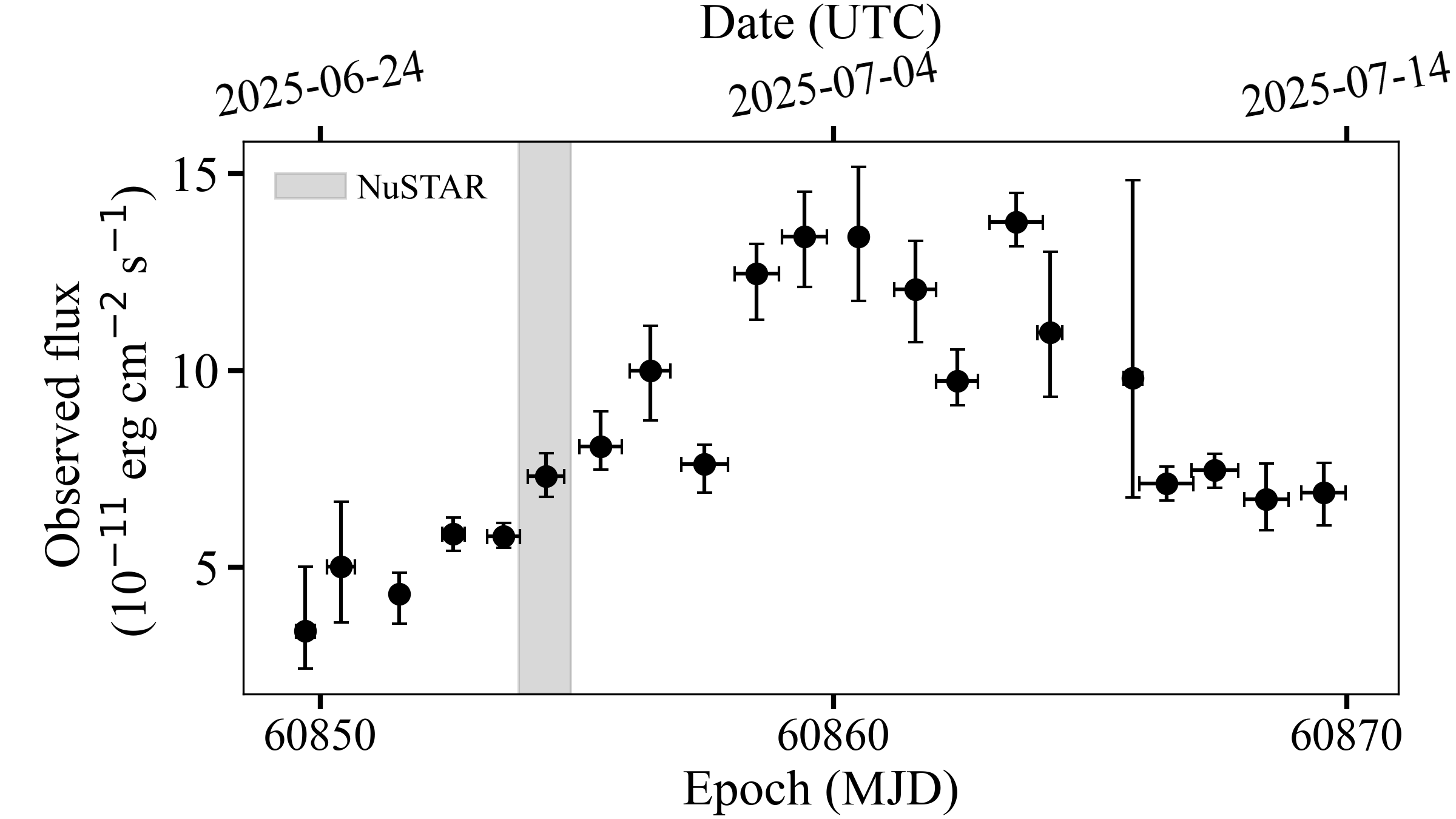}
    \caption{X-ray light curve of \src\ extracted from EP/WXT data collected during the first 21 days of the outburst. Observed fluxes refer to the 0.5--4\,keV energy range and were obtained by combining data from snapshots taken on the same day. The shaded gray area marks the epoch of the \nustar\ observation. }
    \label{fig:wxt_lcurve}
\end{figure}

EP/WXT detected X-ray emission from the position of \src\ in a snapshot beginning on 2025 June 23 at 12:37:24 UTC. From that moment on and until July 13, the source was detected by WXT for a total of 184 times, with individual snapshot exposures ranging from 0.4 to 9.3\,ks and a total exposure time of $\sim\,0.5$\,Ms. Figure~\ref{fig:wxt_lcurve} presents the WXT light curve, binned approximately on a daily timescale, derived by combining all spectral data acquired each day and fitting the resulting spectra with an absorbed power-law model. In the fits, the value for the absorption column density was held fixed to the value obtained from the broadband spectral analysis (Section~\ref{subsubsec:persistent spec}), whereas the power-law photon index, $\Gamma$, was allowed to vary. We note that the WXT light curves occasionally demonstrate suggestive hints consistent with eclipses and X-ray bursts. However, in order to perform a robust timing analysis when only WXT data are available, the limited sensitivity of WXT requires a careful exploration of binning choices and detection criteria over a large dataset, which is beyond the scope of this work. 

\subsubsection{Follow-up X-ray Telescope (FXT)} \label{subsubsec:FXT}
The FXT \citep{chen20} on board EP is a Wolter-I telescope operating in the 0.3--10\,keV band. It has a sensitivity of $\sim10^{-14}\,\flux$ for an exposure of 10\,ks. The energy resolution is 120\,eV at 1.25\,keV (FWHM) and the typical localization uncertainty is $10\arcsec$ (90\% confidence level). EP/FXT consists of two identical modules, FXT-A and FXT-B. During the observation, the two modules can be set in different science modes, including Partial Window mode (PW, with a time resolution of 2\,ms), Full Frame mode (FF, with a time resolution of 50\,ms) and Timing mode (TM, with a time resolution of 23.6\,$\mu$s). 

The first follow-up observation with EP/FXT was performed on 2025 June 25 at 14:09:08 UTC, two days after the WXT detection. In this observation, an X-ray burst and a dip event were detected, which motivated a daily monitoring program with FXT. In this work, we analyzed the dataset of the first 22 EP/FXT observations collected over 21 days following the onset of the outburst. The data were reduced with the \texttt{fxtchain} tool available within the FXT Data Analysis Software Package (\texttt{FXTDAS}; \texttt{v1.20}) following the standard procedure described in the FXT User Guide.\footnote{\url{https://epfxt.ihep.ac.cn/analysis}} Further filtering of the data was performed with \texttt{Xselect}. For PW and FF data, a circular source region with a radius of $60\arcsec$ and an annular background region with inner and outer radii of $120\arcsec$ and $240\arcsec$ were adopted. Moreover, for FF data the inner $10\arcsec$ region around the source position was excluded, in order to mitigate the pile-up effect. For TM data, box regions ($180\arcsec \times 60\arcsec$) aligned with the roll angle of the satellite during each observation were used for both source and background extractions, with the center of the background region shifted from the source location by $360\arcsec$.

\subsection{\nustar}\label{subsec:nustar}
\nustar\ \citep{Harrison2013} is a hard X-ray observatory consisting of two Wolter-I telescopes, focusing X-rays onto two Focal Plane Modules (FPM), labeled A and B. Operating in the 3--79\,keV band, it provides high-sensitivity timing measurements with a time resolution of about 2\,$\mu$s. The energy resolution (FWHM) is 400\,eV at 10\,keV and 900\,eV at 68\,keV. The angular resolution (FWHM) is $18\arcsec$. The sensitivity of \nustar\ is $2\times10^{-15}\,\flux$ in 6--10\,keV and $1\times10^{-14}\,\flux$ in 10--30\,keV, for an exposure of $10^6$\,s. 

A \nustar\ observation targeting \src\ was carried out on 2025 June 27, for a total on-source exposure of 46.7\,ks. We reduced the data from both FPM detectors using standard tools within the \texttt{NuSTARDAS} package (\texttt{v.2.1.4a}). The source region was extracted using a circular region of $100\arcsec$ radius centered at the coordinates of the source. To account for possible spatial non-uniformities in the detector background, we extracted background spectra from four $\sim$50$\arcsec$ circular regions placed away from the source and distributed across the FOV so as to maximize their separation. We extracted 1-s binned light curves with \texttt{Xselect} and inspected them in search for type-I X-ray bursts, eclipses and/or dips. We identified the time intervals corresponding to these events through visual inspection, grouping consecutive time bins that are significantly above (for bursts) or below (for eclipses and dips) the persistent emission level. In order to isolate the persistent emission, we ran \texttt{NuPRODUCTS} to extract spectra and light curves from an event file in which those burst and eclipse/dip intervals had been filtered out. Additionally, separate \texttt{NuPRODUCTS} runs were performed on event files containing only the burst intervals.

\subsection{ULTRACAM}\label{subsec:ultracam}
On 2025 July 21 at 23:39:45 (UTC), we observed \src\ simultaneously in the $u_{\rm s}$, $g_{\rm s}$ and $i_{\rm s}$ filters using the high-speed, triple-beam imager ULTRACAM \citep{dhillon07} on the 3.5-m NTT at La Silla, Chile. We used the so-called Super SDSS filters in ULTRACAM, which are more top-hat in shape than the original SDSS filter set and have significantly higher throughputs, particularly in the $u_{\rm s}$ and $g_{\rm s}$ bands \citep[see][]{dhillon21}. A total of 785 frames were obtained, each of 10.0\,s exposure time. The instrument was used in full-frame mode, no-clear mode, giving a dead time of 0.024\,s between frames, where each ULTRACAM frame is GPS time-stamped to a relative (i.e. frame-to-frame) accuracy of 50\,$\mu$s and an absolute accuracy of 1\,ms \citep{dhillon07}. The source was too faint to detect in the $u_{\rm s}$-band frames and therefore data in this filter will not be discussed further here. 

The ULTRACAM data were reduced using the HiPERCAM data reduction pipeline \citep{dhillon21}. All frames were debiased and then flat-fielded, the latter using the median of twilight-sky frames taken with the telescope spiralling. Due to the crowded field, we used point-spread function (PSF) photometry to extract the counts from \src\ and a number of comparison stars in the same FOV, the latter acting as the reference for the PSF fits, and providing corrections for transparency and extinction variations. The comparison stars were also used for flux calibration via their magnitudes given in the SkyMapper Southern Sky Survey DR4 \citep{onken24}.

\section{Data analysis and results} \label{sec:results}

\subsection{Timing analysis of the clocked bursts}
\label{subsec:burst timing}

\subsubsection{Search for X-ray bursts} \label{subsubsec:burst search}
We searched for X-ray bursts by identifying statistically significant count rate excesses in the data. First, the cleaned, barycentered event data were binned into 2-s light curves. Assuming the background emission is dominated by Poisson noise, we computed for each time bin the chance probability for a noise fluctuation, $p = P(N \ge k \mid \lambda)$, where $k$ is the observed number of counts and $\lambda$ is the mean count rate of the relevant observation. These tail probabilities were converted to Gaussian-equivalent significances by mapping $p$ to a one-sided normal deviate $\sigma_g$ via $p = \tfrac{1}{2}\,[1 - \mathrm{erf}(\sigma_g/\sqrt{2})]$, yielding $\sigma_g = \sqrt{2}\,\mathrm{erf}^{-1}(1 - 2p)$ expressed in units of standard deviations \citep{cowan1998}. 

Burst candidates were defined as time intervals where the significance exceeded 3$\sigma$. The peak time of each burst was defined as the center of the time bin with the maximum significance value. Burst start and end times were determined by tracking the rise and decay phases of the significance profile until the significance either ceased to decrease monotonically or fell below a 1$\sigma$ threshold. We note that this threshold corresponds to a constant Poisson tail probability that is independent of the local count rate or noise level, serving only as a conservative criterion for defining burst boundaries. In addition, we estimated the decay timescale for each burst by fitting the light curve with a Gaussian-rise plus exponential-decay model. 

To maximize photon statistics, we performed the timing analysis of the bursts using the full energy band available for each dataset. We verified that the differences in the derived timing parameters when using the full band versus a fixed common energy range (e.g., 3--10\,keV) are negligible as compared to the time resolution of the light curves.

\begin{figure*}
\centering
\includegraphics[width=\textwidth]{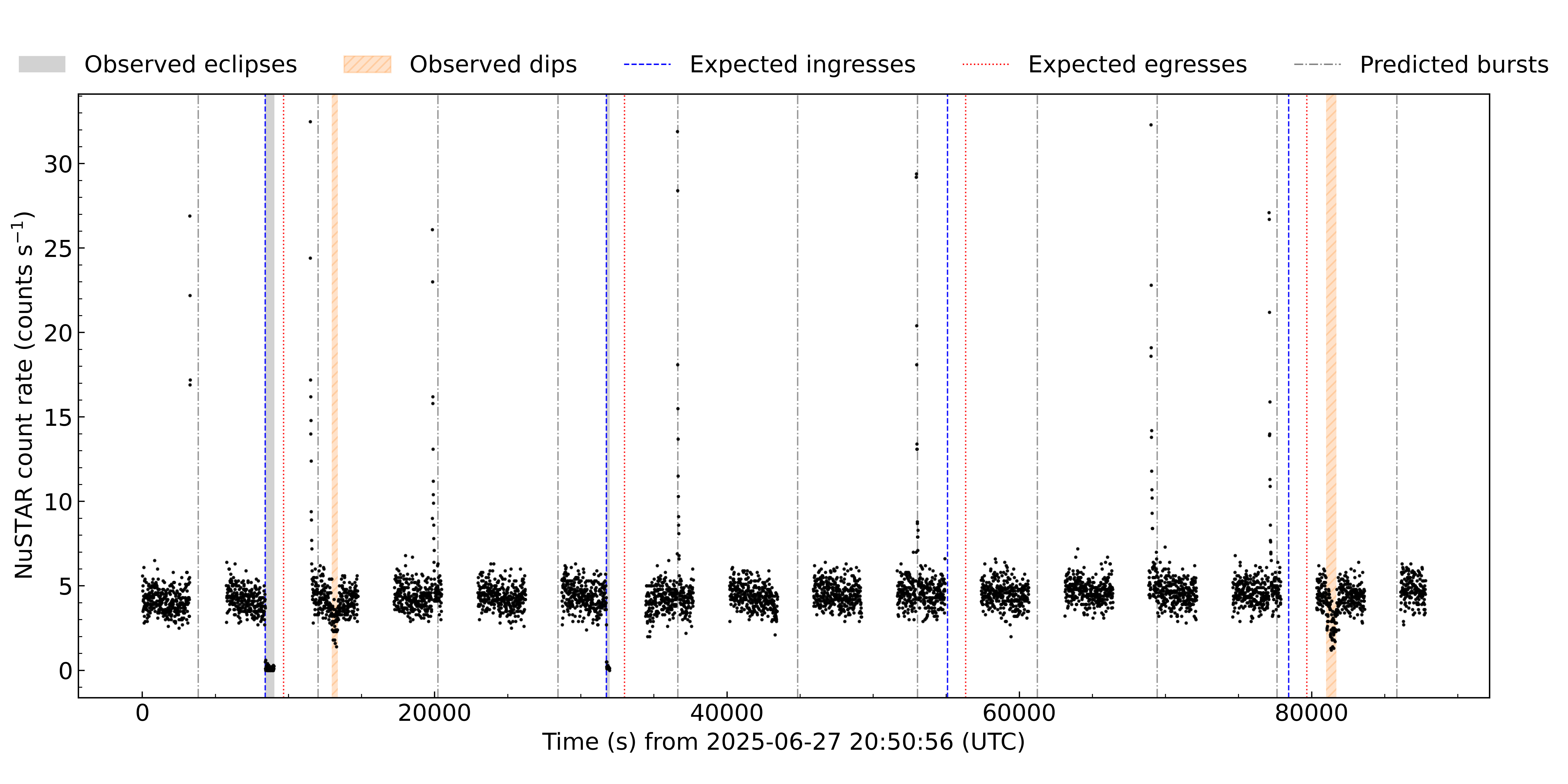}
    \caption{X-ray light curve of \src\ obtained in the \nustar\ observation. The dot-dashed gray lines mark the expected epochs of bursts, using the epoch of the fourth burst as a reference and assuming a constant $t_{\rm rec}$ of 8200\,s. The expected ingress (dashed blue lines) and egress (dotted red lines) epochs of eclipses are calculated from ObsID\,11900304129 (Figure~\ref{fig:fxt19_lc}). The shaded gray areas are the observed eclipse periods. The orange hatched bands are the observed dips. A total of seven bursts, two eclipses and two dips were detected during this observation. }
    \label{fig:nustar_lc}
\end{figure*}

In total, we identified 16 bursts, corresponding to 17 detections: seven in the \nustar\ observation (see Figure~\ref{fig:nustar_lc} for the light curve) and ten in the EP/FXT data (one burst was simultaneously detected by both instruments). Burst properties are listed in Table~\ref{tab:bursts}. 

\subsubsection{Burst recurrence time} \label{subsubsec:burst recurrence}
\begin{figure*}
\centering
\includegraphics[width=\textwidth]{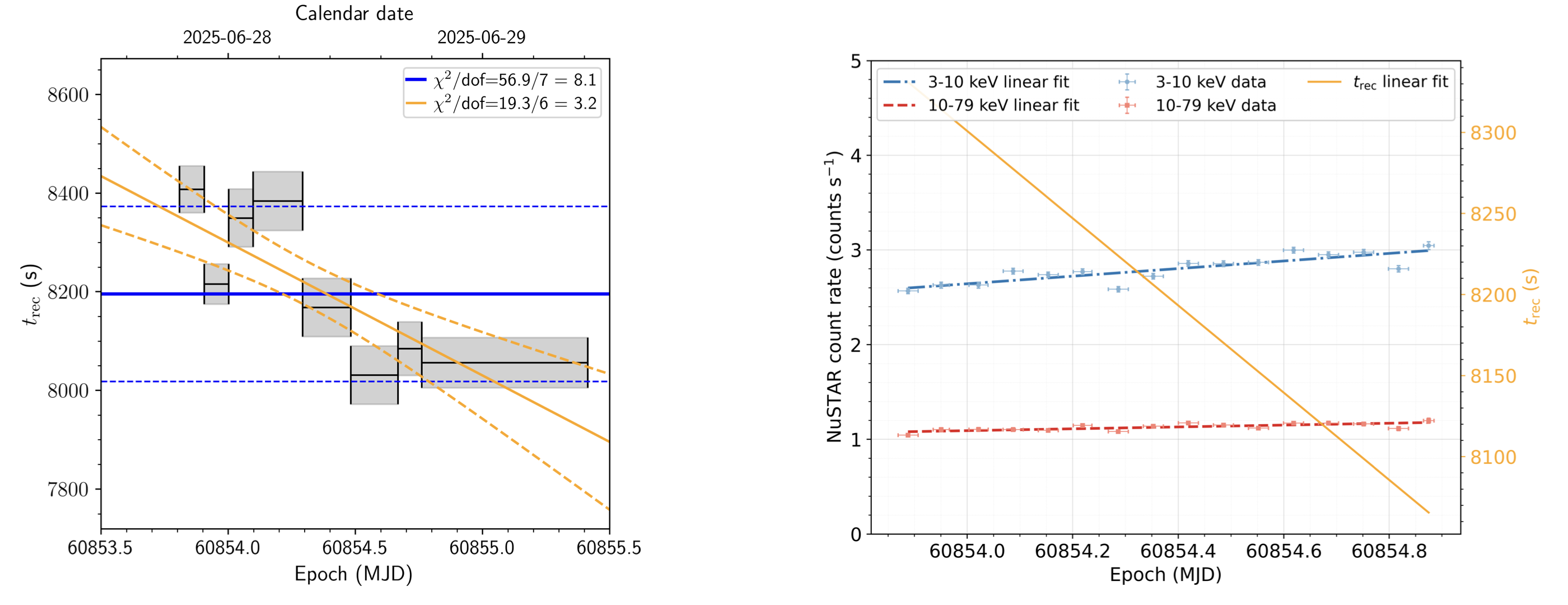}
    \caption{Left: measured recurrence times $t_{\rm rec}$ (black) for the eight consecutive pairs of bursts with unambiguous cycle counts. The grey shaded regions denote the validity range of each measurement, extending from the start of one burst to the start of the subsequent burst. The first burst in the first $t_{\rm rec}$ measurement corresponds to burst No.~3 in Table~\ref{tab:bursts}, while the final burst in the last measurement corresponds to burst No.~11. The blue solid and dashed lines show the weighted median recurrence time, $t_{\rm rec}=8196 \pm 177\,$s, which does not adequately describe the data ($\chi^2_{\rm red}=8.1$). The orange solid and dashed lines represent a linear fit to the $t_{\rm rec}$ measurements, indicating a decreasing trend of $\dot{t}_{\rm rec}=-270\pm110$\,s\,day$^{-1}$, with an improved fit quality ($\chi^2_{\rm red}=3.2$). Right: \nustar\ light curves in 3--10\,keV (blue circles) and 10--79\,keV (red squares) bands and their respective linear fits. The \nustar\ data were binned over individual observational segments to improve the signal-to-noise ratio. The linear fit for $t_{\rm rec}$ (orange solid line) from the left panel is also plotted on top of the light curves. The count rates in both bands show an increasing trend, with best-fit slopes of $0.40\,{\rm cts\,s^{-1}\,day^{-1}}$ (3--10\,keV) and $0.097\,{\rm cts\,s^{-1}\,day^{-1}}$ (10--79\,keV). Quoted errors are at 95\% confidence level. } 
    \label{fig:trec}
\end{figure*}

The recurrence time of the bursts, $t_{\rm rec}$, is defined as the interval between consecutive burst onsets. Due to the intrinsic variability in $t_{\rm rec}$ and the long data gaps in our datasets, we were only able to obtain reliable estimate of $t_{\rm rec}$ for eight consecutive pairs of bursts spanning 1.6\,days around the time of the \nustar\ observation. For these eight pairs of bursts we measured a weighted median $t_{\rm rec}=8196 \pm 177\,$s (95\% confidence level; blue lines in the left panel of Figure~\ref{fig:trec}). Moreover, the data hint at a decreasing trend in $t_{\rm rec}$, and with a linear fit we derived $\dot{t}_{\rm rec} = -270 \pm 110$\,s\,day$^{-1}$ (95\% confidence level; orange lines in the left panel of Figure~\ref{fig:trec}). The details of this analysis are presented in Appendix~\ref{appendix:t_rec}. 

In the type-I X-ray bursters, a decreasing trend in $t_{\rm rec}$ typically points to an increase in the accretion rate and hence in the observed persistent flux (see Section~\ref{subsec:comparison with others} for a more detailed discussion). Motivated by this, we also analyzed the \nustar\ persistent-emission light curves in the 3--10 keV and 10--79 keV bands. As with the $t_{\rm rec}$ analysis, we fitted the data with both a constant and a linear model using maximum likelihood with an intrinsic-scatter term $\sigma_{\rm int}$. In both energy bands, a linear increase is strongly preferred (right panel of Figure~\ref{fig:trec}). For the 10--79 keV band, we obtained, under the linear model, a slope of $0.097\rm\,cts\,s^{-1}\,day^{-1}$ with $\sigma_{\rm int}\simeq 0.018\rm\,cts\,s^{-1}$. A likelihood-ratio test calibrated by 2000 bootstrap simulations gave $p<10^{-3}$, indicating a highly significant increasing trend. Similarly, in the 3--10 keV band we obtained a slope of $0.40\rm\,cts\,s^{-1}\,day^{-1}$ and $\sigma_{\rm int}\simeq 0.071\rm\,cts\,s^{-1}$, with a likelihood-ratio $p<10^{-4}$. In both cases, the intrinsic scatter term makes the residuals compatible with Gaussian noise, and the bootstrap-calibrated likelihood-ratio test confirms the strong preference for a linear increase of the persistent flux during the observation. 

Furthermore, we fitted the relation between $t_{\rm rec}$ and the observed count rate with a power-law model: $t_{\rm rec} = AF^{-\eta}$, where $A$ and $\eta$ are the model parameters to be determined, and $F$ is the \nustar\ count rate during persistent emission. In order to perform this fit, the \nustar\ data in the 3--79\,keV band were first binned over individual observational segments separated by the gaps during the observation. We then paired each $t_{\rm rec}$ value with the average count rate of the segment that is closest in time. Finally, we obtained the following best-fit results (95\% confidence level), confirming the anti-correlation between $t_{\rm rec}$ and the observed persistent flux: $t_{\rm rec} \propto F^{-0.27\pm0.16}$, $\chi^2_{\rm red} = 3.0$ with 6 degrees of freedom (dof). The fitting results are also presented in Figure~\ref{fig:trec_flux_fit}.

\subsection{Timing analysis of eclipses and dips in X-rays} \label{subsec:eclipse&dip}
Apart from the type-I X-ray bursts, the light curves of \src\ obtained by both EP/FXT and \nustar\ clearly demonstrated two more types of prominent variabilities, eclipses and dips. 

We first investigated the periodicity of the eclipses by applying the Bayesian block algorithm \citep{Scargle2013} to photon arrival times in each observation. This method identifies statistically significant changes in the source count rate and divides a light curve into several segments (blocks) based on these changes. The count rate is therefore roughly constant within each block yet drastically changes at the edges between neighboring blocks. This algorithm is well-suited to detecting the sharp transitions expected during eclipse events. Eclipses were identified as intervals during which the count rate dropped to the background level. The start of each eclipse (ingress) was defined as the beginning of the block immediately following a sharp decline in count rate, while the end (egress) was defined as the termination of the block preceding a sharp recovery to the out-of-eclipse level. The statistical uncertainty on each ingress and egress time was conservatively estimated as half the interval between the last photon before and the first photon after the identified block edge.

\begin{figure}
\includegraphics[width=0.48\textwidth]{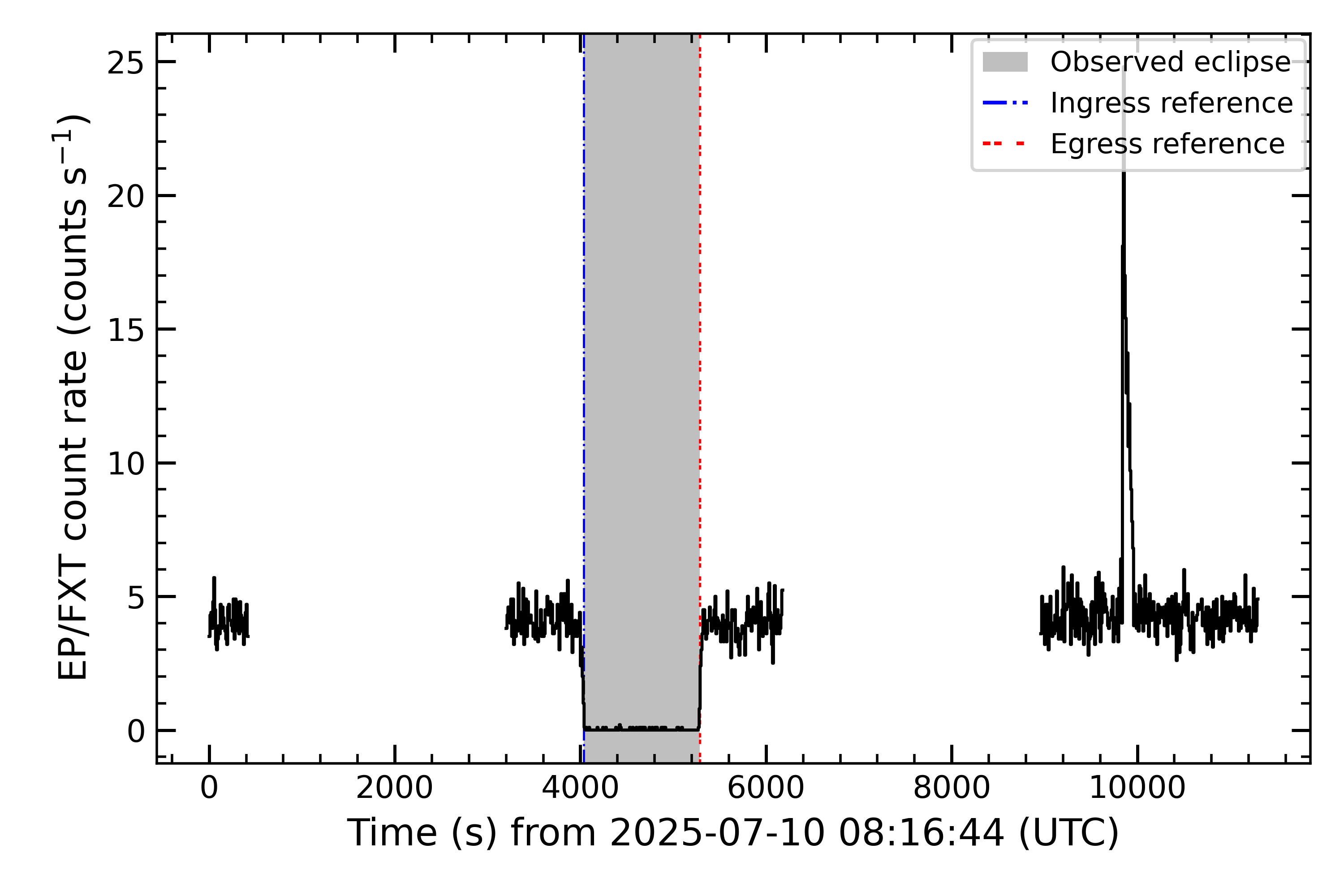}
    \caption{X-ray light curve of \src\ from ObsID\,11900304129. This is the only EP/FXT observation where an entire eclipse event was captured. We use the ingress and egress epochs observed in this eclipse event as reference points to calculate the expected ingress and egress epochs in other observations. See Figure~\ref{fig:all_fxt_lc} for all the FXT light curves.}
    \label{fig:fxt19_lc}
\end{figure}

Only one complete eclipse was observed in our dataset (EP/FXT, ObsID\,11900304129; see Figure~\ref{fig:fxt19_lc}), while the remaining events were either ingress-only or egress-only detections. This was primarily due to the limited duration of the good time intervals, which often started or ended during the eclipse itself. In total, we identified seven ingress and two egress times across six distinct eclipses. 

To assign integer cycle numbers to each event, we used a provisional orbital period estimated from the separation between consecutive ingress times observed with \nustar\ (Figure~\ref{fig:nustar_lc}). We also adopted a provisional eclipse duration of $D_{\rm obs} \approx 1260$\,s, measured from the complete eclipse, to provide a consistent phase reference for all eclipses, including the partial ones.

The mid-eclipse time of the n$^{\rm th}$ orbital cycle ($M_n$) is defined as $M_n = T_0 + nP_{\rm orb}$, where $T_0$ is the reference mid-eclipse epoch for cycle zero, and $P_{\rm orb}$ is the orbital period. The ingress and egress times are modeled as 
$
    T^{(n)}_{\text{ingress}} = M_n - \frac{D_n}{2}, \quad
    T^{(n)}_{\text{egress}} = M_n + \frac{D_n}{2},
$
where $D_n$ is the eclipse duration for cycle $n$. Rather than assuming a fixed eclipse duration, we model each $D_n$ as a random draw from a parent Gaussian distribution with mean $D_{\star,X}$, representing the typical eclipse duration, and standard deviation $\sigma_D$, which captures intrinsic variability from cycle to cycle. The joint posterior over the model parameters was sampled using Markov Chain Monte Carlo \citep[MCMC;][]{Hastings1970} techniques. We measured $P_{\rm orb}=6.48301 \pm 0.00003$\,hr, $T_0 = 60852.62254(6)$\,MJD, $D_{\star,X} = 1245.5^{+6.9}_{-6.5}$\,s and $\sigma_D = 14.0$\,s (95\% upper limit: 68.6\,s). Uncertainties represent the 16th and 84th percentiles of the marginal posterior distributions.

\begin{figure*}
\includegraphics[width=\textwidth]{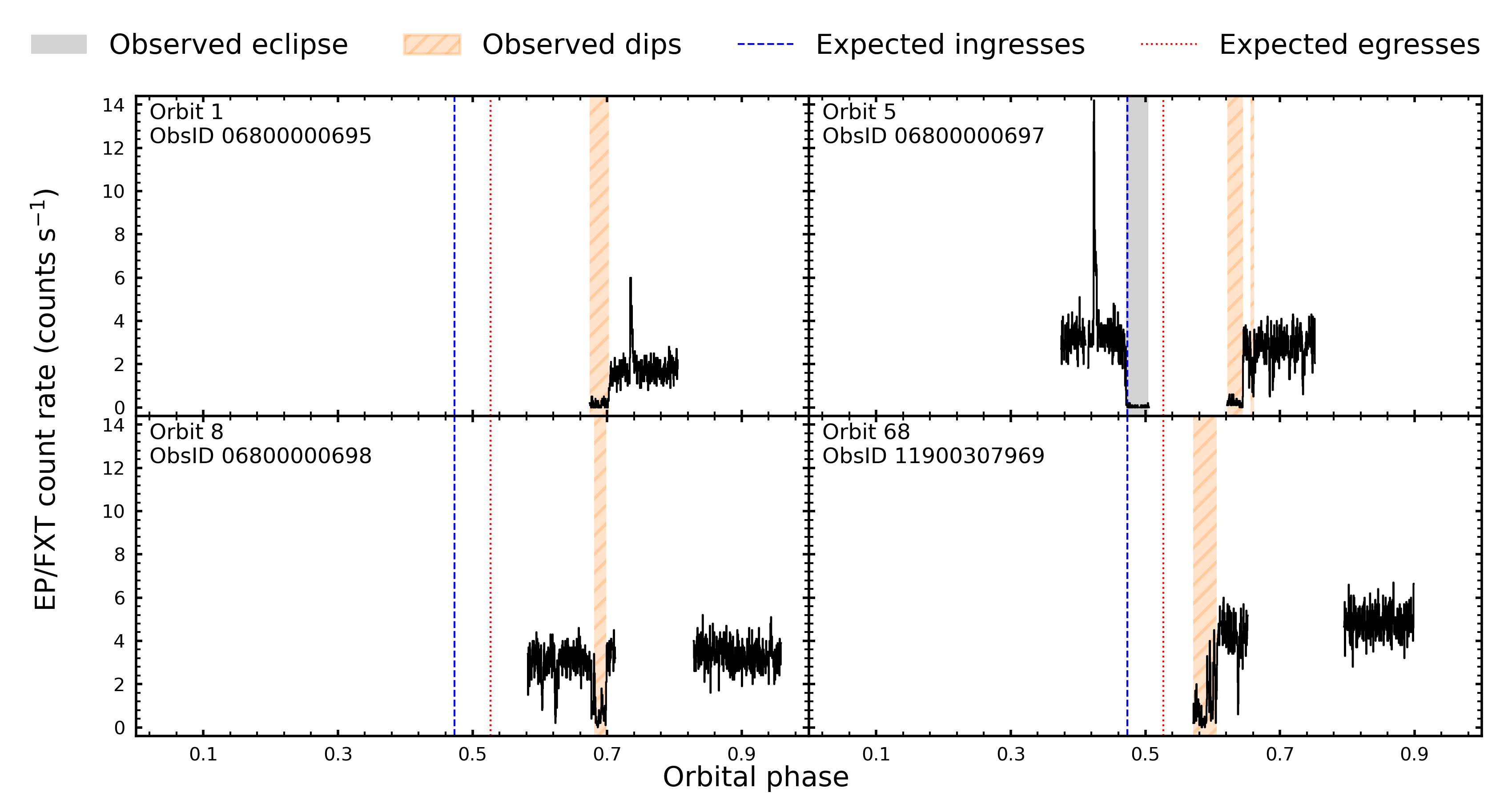}
    \caption{Phase-resolved light curves of four EP/FXT observations with dips (orange hatched bands). Each panel shows the data obtained over one orbit of the binary system. The orbit observed by the first FXT observation is defined as Orbit 1. The expected ingress (dashed blue lines) and egress (dotted red lines) epochs are calculated from ObsID\,11900304129. The shaded gray area is the observed eclipse period. See Figure~\ref{fig:all_fxt_lc} for all the FXT light curves.}
    \label{fig:fxt_dips}
\end{figure*}

Apart from the eclipses, we also identified a total of six dip events from our main dataset with the Bayesian block algorithm: two in the \nustar\ observation (Figure~\ref{fig:nustar_lc}) and four in the EP/FXT dataset (Figure~\ref{fig:fxt_dips}). Similar to eclipses, dips also involve significant rises and drops in the count rate, but they were distinguished from eclipses because of their irregular recurrence and variable duration. Moreover, during the dip events, the count rate never dropped to the background level as was the case for eclipses, and the light curve demonstrated irregular short-term variations that never occurred in eclipses. These differences point to fundamentally different physical mechanisms behind these two types of events. We note that, in order to mitigate the impact of random short-term statistical fluctuations, we conservatively retained only the detected dip candidates which lasted for over 100\,s. 

\subsection{Searches for coherent pulsations and burst oscillations}
\label{subsec:coherent signal}
We used the pulsar timing software package \texttt{PRESTO} \citep{Ransom2011} to perform Fourier-domain accelerated searches for coherent signals in the data collected during the persistent emission periods by both EP/FXT and \nustar. Data were divided into chunks with a length of 2.3 ks, corresponding to roughly one tenth of the orbital period estimated from the periodic recurrence of the eclipses. We chose this segment duration to maximize sensitivity (proportional to $\sqrt{T_{\rm obs}}$, where $T_{\rm obs}$ is the duration of the observation) while avoiding the loss of power caused by non-linear acceleration Doppler smearing, known to be significant for $T_{\rm obs}$ higher than $\sim P_{\rm orb}/10$ \citep{Ransom2001, Ransom2002, Andersen2018}. No statistically significant signals were found. The best constraint on the pulsed-fraction upper limit was obtained with ObsID\,06800000720, which is 11\% (99\% confidence level, in 0.5--10\,keV band). We also performed the same analysis within 0.5--2\,keV and 2--10\,keV bands in order to separate the emissions from disk and corona, and the derived pulsed-fraction upper limits are 13\% (ObsID\,06800000724) and 22\% (ObsID\,06800000720), respectively. 

We also searched for burst oscillations in each type-I X-ray burst using sliding Fourier windows of 1, 2, and 4\,s duration. For each burst, the search covered the interval from 10\,s before the burst onset to the end of the burst, corresponding to a total duration of approximately 130\,s. Power spectra were computed in the frequency range of 50--2000\,Hz using Leahy normalization. Adjacent time windows were stepped by half of the window length. The total number of independent trials per burst was therefore $\sim 1.4 \times 10^{6}$, accounting for the number of time windows, frequency bins, and window lengths. No statistically significant oscillation signal was detected after correcting for the number of trials. We derived upper limits on the fractional rms amplitude by converting the detection-threshold power to rms amplitude using $r_{\rm rms} = \sqrt{P_{\rm th}/2N}$, where $N$ is the total number of photons in each time window. For a global $3\sigma$ confidence level, this corresponds to a Leahy-normalized power threshold of $P_{\rm th} \approx 40$. 

This analysis was performed for the eight bursts observed by EP/FXT in TM, and the resulting $3\sigma$ upper limits on the oscillation amplitude depend on the burst brightness and integration time. The most constraining limits we obtained are approximately $65-84\%$ rms for 1-s windows, $49-68\%$ rms for 2-s windows, and $36-50\%$ rms for 4-s windows, with the tightest constraints obtained from the brightest bursts and the longest integration times. For \nustar\ bursts, with the same analysis, the limits we derived are $49-58\%$ rms for 1-s windows, $35-41\%$ rms for 2-s windows, and $27-31\%$ rms for 4-s windows.

\subsection{X-ray spectral analysis} \label{subsec:spectra}

\subsubsection{Persistent emission} \label{subsubsec:persistent spec}
In this section, we report on the spectral analysis of the persistent emission from \src\, excluding bursts, eclipses and dips. All EP/FXT and \nustar\ spectra were grouped using the optimal binning following \citet{Kaastra2016}, while maintaining at least 25 counts per bin to allow the use of $\chi^2$ statistics. Spectral fits were performed with \texttt{Xspec} v12.12.1. In all cases, we applied the \texttt{TBabs} model to account for interstellar absorption, adopting photoelectric cross-sections and elemental abundances from \citet{Verner1996} and \citet{Wilms2000}, respectively. When combining data from multiple instruments, a \texttt{constant} component was added to account for cross-calibration uncertainties, and we verified that differences in the constant values between instruments remained within 20\%. 

\begin{figure}
\includegraphics[width=0.48\textwidth]{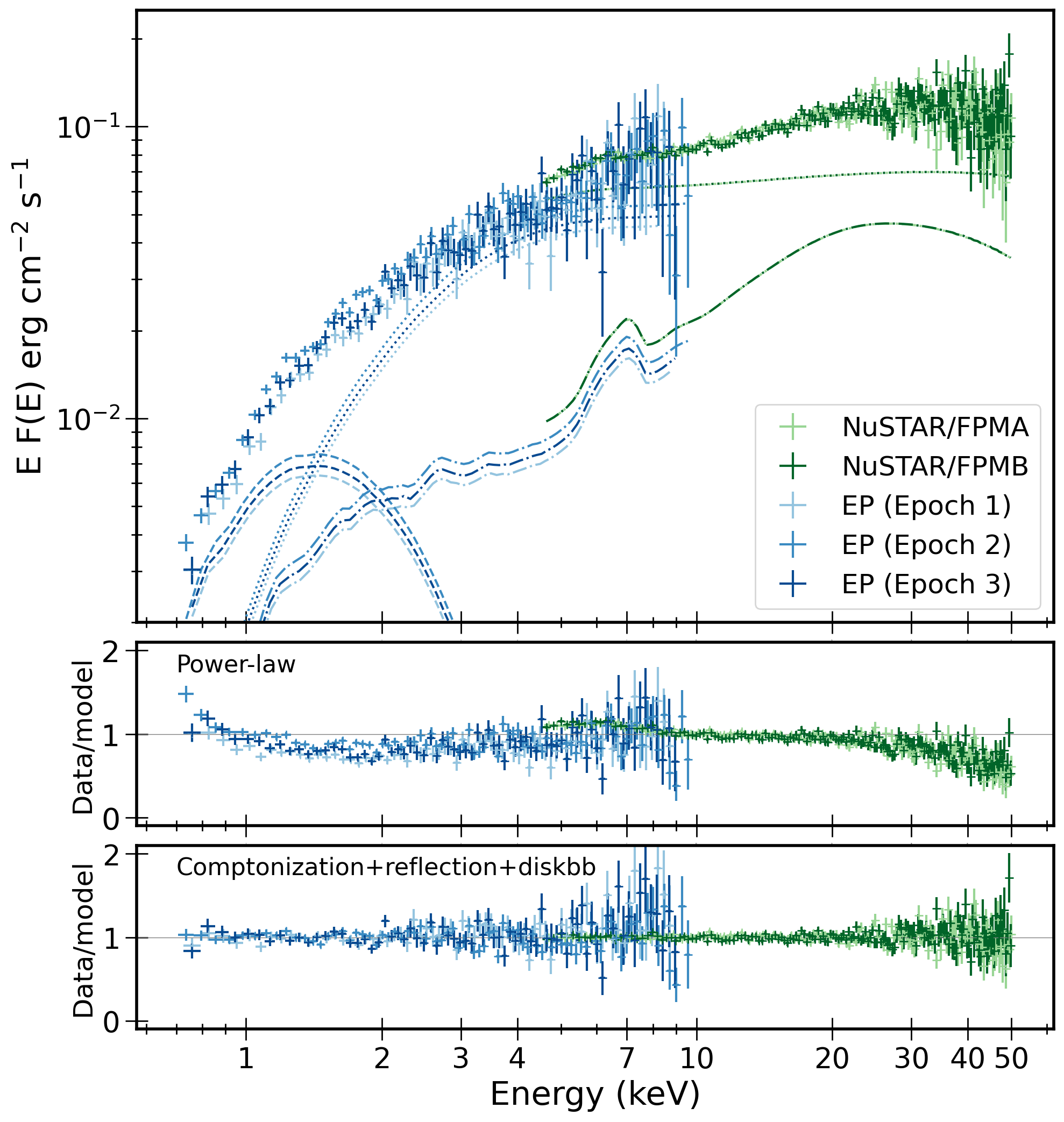}
    \caption{Broadband EP/FXT (blue) and \nustar\ (green) spectra with the best-fit Model 1 (top panel), and residuals from Model 0 (middle panel) and Model 1 (bottom panel). The EP/FXT data are from ObsIDs\,0680000699 (Epoch 1), 0680000706 (Epoch 2) and 0680000707 (Epoch 3). Different line styles were adopted to distinguish different components: dot for \texttt{diskbb}, dash for \texttt{thcomp}$\times$\texttt{bbodyrad} and dash-dot for \texttt{relxillCp}. }
    \label{fig:broadband-spec}
\end{figure}

With the aim to perform a broadband spectral analysis, we paired our \nustar\ spectrum with all the EP observations performed within the \nustar\ observing window, i.e., ObsIDs\,0680000699, 0680000706 and 0680000707 (see Table~\ref{tab:log} for more details). An initial exploratory fit with a simple \texttt{TBabs}$\times$\texttt{powerlaw} model (Model 0) was found unacceptable, as significant unmodeled residuals were present below 2\,keV, around 7\,keV and beyond 20\,keV (see Figure~\ref{fig:broadband-spec}). The residuals indicate the presence of both a soft thermal component and reflection features. We therefore adopted a more physically motivated model, assuming the canonical three-component model (including blackbody emission from the NS/boundary layer, multicolor disk blackbody emission from the disk and Compton-scattering from a hot inner flow or corona) for NS-LMXBs spectra by \cite{LinRemillardHoman2007}. In practice, the \texttt{powerlaw} component was replaced with a \texttt{bbodyrad} convolved with \texttt{thcomp} \citep{Zdziarski2020_thcomp}, to jointly model the Comptonization and the seed photon spectra. This component assumes a geometry where the photons coming from the NS or the boundary layer around it are Compton-scattered by the corona. To represent the soft thermal emission, we added a \texttt{diskbb} component. As required by the observed residuals, we also included a component to model the contribution from the reflection spectrum arising from the reprocessing by the disk of the incident Comptonization spectrum emitted by the corona \citep[see, e.g.][for a recent review on reflection in NS-LMXBs]{Ludlam2024}. Such a component was accounted for using the self-consistent \texttt{relxillCp} model \citep{Garcia2014}. To capture the low-energy roll-over in the seed photon spectrum --- unaccounted for in \texttt{relxillCp}, which assumes a fixed and unrealistically low seed temperature of 0.05\,keV --- we included an \texttt{expabs} multiplicative component, characterized by the cut-off energy $E_{\rm cut-off, low}$ parameter. By running \texttt{ftest}, both \texttt{diskbb} and \texttt{relxillCp} were found to be highly statistically significant, with probabilities of improvement by chance of $\sim$10$^{-20}$ and $\sim$10$^{-29}$, respectively. Despite the inclusion of reflection, residual absorption features remained near $\sim$7\,keV, prompting us to add a \texttt{gaussian} absorption line, a common feature in eclipsing/dipping LMXBs \citep[for a review, see][]{DiazTrigo2016}. Our final model reads therefore: 
\begin{align}
\text{Model 1:\ } \texttt{TBabs} \times ( \texttt{thComp} \times \texttt{bbodyrad} 
+ \texttt{diskbb} \notag \\
+ \texttt{expabs} \times \texttt{relxillcp} 
+ \texttt{gaussian} ) \notag
\end{align}
This continuum modeling approach is commonly adopted in broadband spectral studies of NS-LMXBs, as demonstrated by several recent works \citep{Gambino2019,Mondal2020,Marino2023,Banerjee2024,Marino2026}. An overview of all the parameters included in the model is reported in Table~\ref{tab:fit_broadband}. For physical consistency, we tied $E_{\rm cut-off, low}$ to $3\times kT_{\rm bb}$, the photon indices $\Gamma$ and electron temperatures $kT_{\rm e}$ of \texttt{thcomp} and \texttt{relxillcp} and fixed the reflection fraction parameter $f_{\rm refl}$ to $-1$ to only consider the reflection spectral component. Some parameters were not well constrained by the fit, so that we froze them to typically adopted values for LMXBs \citep[e.g.][]{Marino2022, Ludlam2024, Malacaria2025, LaMonaca2025}, i.e., the emissivity index $\epsilon$ to 3, the Fe abundance A$_{\rm Fe}$ to 1, and the density parameter $\log{N}$ to 20. For the binary inclination $i$, which was also not well constrained by the fit due to degeneracies, we fixed the value to 74$^\circ$ (see Section~\ref{subsec:binary properties} for the discussion on the binary inclination $i$). 

\begin{table*}
\centering
\caption{Results of the broadband spectral analysis for the persistent emission. Quoted errors reflect 90\% confidence level. The parameters that were held constant during the fits are reported between round parentheses. $R_{\rm g}$ represents the gravitational radius. The reported flux values correspond to the 1--10\,keV energy range. }
\begin{tabular}{ l l l l l }
\toprule
Component & Parameter & Unit & Description & Value \\
\midrule
\texttt{TBabs} & $N_{\rm H}$ & 10$^{22}$ cm$^{-2}$ & Equivalent hydrogen column density & 0.48$^{+0.08}_{-0.07}$ \\
\multirow[t]{3}{*}{\texttt{thComp}} & $\Gamma$ &  & Power-law index of the Comptonization spectrum & 1.90$^{+0.04}_{-0.03}$ \\ 
& $kT_e$ & keV & Electron temperature of the corona & 22$^{+20}_{-3}$ \\
& $f_{\rm cov}$ &  & Covering fraction & (1.0) \\
\multirow[t]{2}{*}{\texttt{bbodyrad}} & $kT_{\rm bb}$ & keV & Blackbody temperature & 0.80$^{+0.30}_{-0.08}$ \\
& $K_{\rm bb}$ &  & Blackbody normalization & 28$^{+11}_{-17}$ \\
\multirow[t]{2}{*}{\texttt{diskbb}} & $kT_{\rm disk}$ & keV & Inner disk temperature & 0.30$^{+0.25}_{-0.05}$ \\
& $K_{\rm disk}$ &  & Disk normalization & 400$^{+890}_{-280}$ \\ 
\texttt{expabs} & $E_{\rm cut-off, low}$ & keV & Low energy cut-off & $=3\times kT_{\rm bb}$ \\ 
\multirow[t]{12}{*}{\texttt{relxillCp}} & $i$ & $^\circ$ & Binary inclination & (74.0) \\ 
& $a*$ &  & Spin parameter & (0) \\ 
& $R_{\rm in}$ & $R_{\rm g}$ & Inner disk radius & 38$^{+14}_{-9}$ \\ 
& $R_{\rm out}$ & $R_{\rm g}$ & Outer disk radius & (1000) \\
& $\epsilon$ &  & Disk emissivity  & (3.0) \\
& $z$ &  & Redshift to the source & (0) \\
& $\Gamma_{\rm relxill}$ &  & Power-law index of the incident spectrum & $=\Gamma$ \\ 
& $\log{\xi}$ &  & Disk ionization & 2.13$^{+0.10}_{-0.09}$ \\
& $\log{N}$ & cm$^{-3}$ & Disk density & (20.0) \\
& A$_{\rm Fe}$ &  & Fe abundance of reflecting material & (1.0) \\
& $kT_{\rm e, refl}$ & keV & Electron temperature of the corona & $=kT_{\rm e}$ \\
& $f_{\rm refl}$ &  & Reflection fraction & ($-1.0$) \\ 
& $K_{\rm refl}$ &  & Reflection normalization & 0.0021$^{+0.0004}_{-0.0008}$ \\
\multirow[t]{3}{*}{\texttt{gaussian}} & $E_{\rm line}$ & keV & Line centroid energy & (6.97) \\
& $\sigma_{\rm line}$ & keV & Line width & (0.2) \\ 
& $K_{\rm line}$ &  & Line normalization & $-(6.0\pm2.0)\times10^{-5}$ \\
\texttt{cflux} & $F_{1-10 \ {\rm keV}}$ & ($\times$10$^{-10}$) \flux & X-ray unabsorbed flux & 2.070$\pm$0.008\\
& $\chi^2/\nu$ &  & $\chi^2$ for $\nu$ degrees of freedom & $544/526$ \\  
\bottomrule
\end{tabular}
\label{tab:fit_broadband}
\end{table*}

\begin{figure}
\includegraphics[width=0.48\textwidth]{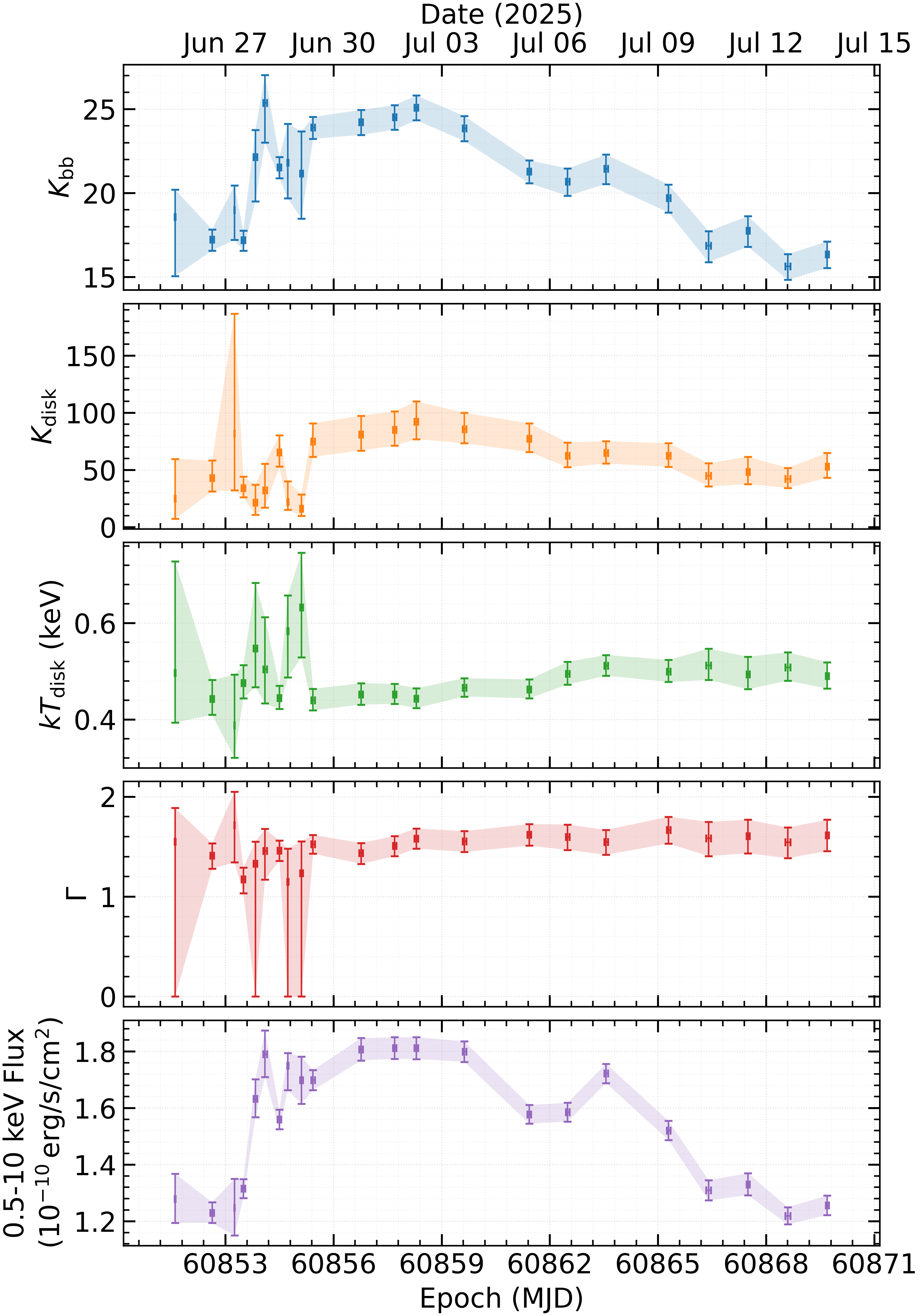}
    \caption{Evolution of key spectral parameters from the 22 FXT observations presented in this work. The model adopted for spectral analysis is $\texttt{TBabs} \times (\texttt{thComp} \times \texttt{bbodyrad} + \texttt{diskbb})$. The reported flux is the observed flux. Shaded areas denote uncertainties at 90\% confidence level. }
    \label{fig:fxt_spec_evolution}
\end{figure}

In order to investigate the long-term evolution of the persistent emission, we performed similar spectral analysis for the rest of the FXT observations as well. Due to a lack of high-energy coverage above 10\,keV and the absence of clear residuals due to reflection when only FXT data are available, we removed \texttt{relxillCp} from the spectral model, thus adopting a simplified version that reads: $\texttt{TBabs} \times (\texttt{thComp} \times \texttt{bbodyrad} + \texttt{diskbb})$. To better constrain the parameters, during spectral fitting, we fixed the values of the hydrogen column density ($N_{\rm H}$), electron temperature of the corona ($kT_{\rm e}$) and blackbody temperature ($kT_{\rm bb}$) to the values derived from the joint fitting with \nustar\ (see Table~\ref{tab:fit_broadband}), and the value of covering fraction ($f_{\rm cov}$) to $1$. Data collected by both FXT-A and FXT-B were used for spectral analysis, except that for TM we ignored the data below $1\,{\rm keV}$ and the data above $9\,{\rm keV}$ due to known calibration issues (private communication). The unfolded EP/FXT spectra and the spectral fitting residuals are shown in Figure~\ref{fig:all_fxt_spec}. The evolution of the key spectral parameters is shown in Figure~\ref{fig:fxt_spec_evolution} and a detailed discussion is given in Section~\ref{subsec:spec discussion}. 

\subsubsection{Type-I X-ray burst emission} \label{subsubsec:burst spec}
We performed time-resolved spectral analysis for all the bursts. For each burst, we divided the whole time span into several segments and performed spectral fitting for each segment. For the seven bursts in the \nustar\ dataset (Table~\ref{tab:bursts}), the \texttt{Xspec} model adopted in the analysis is $\texttt{TBabs} \times (\texttt{nthComp}+\texttt{bbodyrad})$, where $\texttt{nthComp}$ is used to account for the persistent emission during the burst and $\texttt{bbodyrad}$ models the burst spectrum. During the spectral fitting, the values of $N_{\rm H}$, $\Gamma$, $kT_{\rm e}$, $kT_{\rm bb}$ (note that $kT_{\rm bb}$ here is a parameter of $\texttt{nthComp}$, representing the temperature of the seed photons for Comptonization) were fixed to those derived from the broadband spectral analysis of the persistent emission (Table~\ref{tab:fit_broadband}). We further tied the normalization of $\texttt{nthComp}$, $K_{\rm Comp}$, across the different segments in order to better constrain the variations of the $\texttt{bbodyrad}$ model parameters used to describe the thermal emission during the bursts, given the relatively low count statistics in each segment. A similar analysis was also performed for bursts in the EP/FXT dataset (Table~\ref{tab:bursts}), adopting a simplified model of $\texttt{TBabs} \times \texttt{bbodyrad}$ due to the limited bandpass of FXT. In order to account for the persistent emission which is not included in such a model, we used the pre-burst persistent spectrum as background. We were able to derive acceptable fits (with $\chi^2_{\rm red} < 1.5$) and constrain the spectral parameters for most of the time segments. The evolution of $F_{\rm bst,bol}$, $kT_{\rm bst,bb}$ and $K_{\rm bst,bb}$ for each burst is presented in Figure~\ref{fig:NuSTAR_burst_time_res} (\nustar) and Figure~\ref{fig:FXT_burst_time_res} (EP/FXT). We found no evidence for photospheric radius expansion \citep[see e.g.][]{Lewin1993} in either of the datasets, implying that the burst luminosities are sub-Eddington. The peak burst temperature can reach $\sim2$\,keV (e.g. burst No.~7) and the peak bolometric flux during the bursts is $\sim2.6\times10^{-9}\,\flux$ (burst No.~16).

Burst No.~8 was simultaneously detected by both \nustar\ and EP/FXT, providing an excellent opportunity to investigate its spectral evolution over a broad energy range. We performed time-resolved spectral analysis from the onset to the end of the burst using time bins of 5\,s, 10\,s, 20\,s, and 50\,s, combining data from both instruments. We adopted two different methods to analyze the burst spectra. The first is the standard approach, commonly referred to as the ``classical'' method, which assumes that the persistent emission remains unchanged before and during the burst. This assumption is also made for the other 15 bursts. In this method, the 100\,s pre-burst spectrum was used as the background in the spectral fitting and the burst emission was modeled with $\texttt{TBabs} \times \texttt{bbodyrad}$ in \texttt{XSPEC}. The second method is the so-called variable-persistent emission $f_a$ method \citep{Worpel2013}, which allows the persistent emission to vary in intensity during the burst while assuming that its spectral shape remains unchanged. We adopted the $\texttt{TBabs} \times (\texttt{nthComp}+\texttt{bbodyrad})$ model and allowed only $K_{\rm Comp}$ to vary, thereby accounting for changes in the persistent emission during the burst. 

The spectral analysis results are presented in Figure~\ref{fig:burst8}. At the onset of burst No.~8, the bolometric flux rises rapidly to a peak of $\sim(2-3)\times10^{-9}\,{\rm erg\,cm^{-2}\,s^{-1}}$, accompanied by a high blackbody temperature of $kT_{\rm bst,bb}\approx 2.0-2.2$ keV. During the subsequent decay phase, the flux decreases monotonically, while $kT_{\rm bst,bb}$ shows a gradual cooling trend toward $\sim 1.5-1.7$ keV. In contrast, $K_{\rm bst,bb}$ initially increases around the burst peak and then declines, consistent with the cooling tail behavior commonly observed in thermonuclear bursts. The results obtained with the classical and $f_a$ methods are broadly consistent in terms of the overall flux and temperature evolution. However, the $f_a$ method indicates a significant enhancement of the persistent emission during the early phase of the burst, as reflected by an increase in $K_{\rm Comp}$ that peaks near the burst maximum and subsequently decreases toward the pre-burst level. This suggests that the burst temporarily amplifies the persistent emission component, while the intrinsic burst emission follows a standard cooling evolution trend. 

\subsubsection{Ratio of accretion energy to thermonuclear energy} \label{subsubsec: alpha}
The above spectral-timing analysis allowed us to evaluate another key parameter, the ratio between the accretion energy released in the interval between two consecutive bursts and the thermonuclear energy released during the second burst ($\alpha$). Among the 16 bursts detected, there are only three pairs of bursts that were consecutive, i.e., no more bursts likely occurred in between. Specifically, the three pairs are bursts No.~4--5, No.~5--6 and No.~9--10, which allowed us to estimate $\alpha$ for bursts No.~5, 6 and 10. For each of the three pairs, we performed spectral fitting both for the burst and for the persistent emission between the bursts using the same methods described in Section~\ref{subsubsec:persistent spec} and Section~\ref{subsubsec:burst spec}, and we estimated the bolometric fluxes by extrapolating the best-fit models to the 0.01--100\,keV band. $\alpha$ was then derived as the ratio between the bolometric fluence of the persistent emission and that of the burst. We estimated $\alpha \approx130$ for bursts No.~5 and 6, and $\alpha \approx120$ for burst No.~10. The implication of these values is discussed in Section~\ref{subsec:comparison with others}.

\subsection{The optical counterpart} \label{subsec:optical counterpart}
\begin{figure}
\centering
\includegraphics[width=0.4\textwidth]{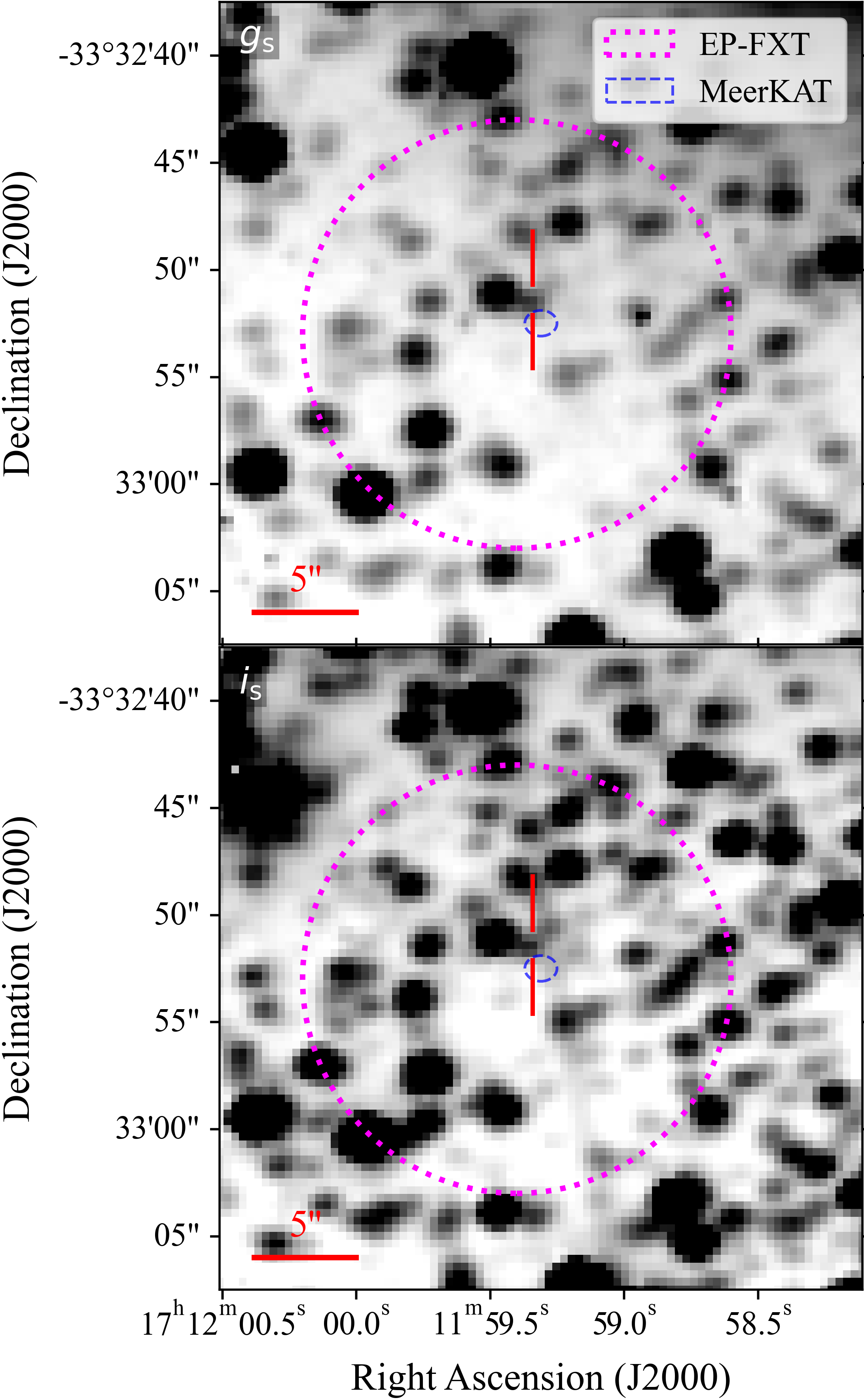}
    \caption{A deep stack of the 2.2 hours of optical data obtained with ULTRACAM in the $g_{\rm s}$ (top) and $i_{\rm s}$ (bottom) bands. The 785 images were individually shifted to account for image motion prior to adding together. The images shown are $30\arcsec$ on a side. North is up and East to the left. The optical counterpart is indicated with the red tick marks. The positional error regions of the X-ray source (EP/FXT) and the radio counterpart (\emph{MeerKAT}) are overlaid. }
    \label{fig:optical image}
\end{figure}

We constructed light curves for all of the stars within the various X-ray, radio and optical error circles reported in Section~\ref{sec:intro}. Only one target (see Figure~\ref{fig:optical image} for the optical image) showed any significant variability. Moreover, this variability is in the form of an optical eclipse that coincides with the time of the X-ray eclipse --- further discussed in Section~\ref{subsubsec:optical eclipse} --- giving us great confidence that we have indeed identified the optical counterpart. 

\subsubsection{The optical eclipse} \label{subsubsec:optical eclipse}
Using the simultaneous ULTRACAM and EP observations, we studied the eclipse duration over different energy bands. Figure~\ref{fig:optical light curves} shows that the optical flux exhibits a smooth transition during the eclipses, in stark contrast to the sharp behavior observed in the X-rays. In order to model specifically these differences, we fit each optical light curve with a piecewise-linear trapezoidal model (after removing the flare, see Section~\ref{subsubsec:optical flare}) that includes seven free parameters: the out-of-eclipse flux levels (allowed to differ before and after the eclipse), the in-eclipse flux, the ingress start time, and the durations of the three segments. The profile of the trapezoidal model is flat outside the eclipse, linear during ingress and egress, and constant during totality. Fits were performed over a time interval fully covering the eclipse and displaying stable out-of-eclipse emission, by means of a non-linear least-squares algorithm. Parameter uncertainties were derived from 300 bootstrap realizations per band, where synthetic light curves were generated by perturbing the best-fit model with Gaussian noise and refitting them.

The analysis was repeated using several bin sizes between 50 and 100\,s, given that the parameter values may depend to some extent on the temporal binning of the light curves. Across all tested binnings, the total eclipse duration ranges from $D_{\star,g}\approx1370–1610$\,s in the $g_{\rm s}$ band to $D_{\star,i}\approx1570–2015$\,s in the $i_{\rm s}$ band. Thus, the $i_{\rm s}$-band eclipse is consistently longer than in the $g_{\rm s}$-band by $\approx200$--400\,s, with a statistical significance of $\gtrsim3\sigma$ across all tests. We also measured the eclipse depth, defined as the difference between the average out-of-eclipse flux and the flux during totality, expressed as a fraction of the out-of-eclipse level. Across all sampled binnings, the eclipse removes $\approx50\%$ of the $g_{\rm s}$-band flux and $\approx35\%$ of the $i_{\rm s}$-band flux.

Combined with the X-ray eclipse results (see Section~\ref{subsec:eclipse&dip}), our measurements show that the eclipse is shorter and removes a larger fraction of the emission at shorter wavelengths, revealing a clear wavelength dependence in the occulted component.

\subsubsection{The optical flare} \label{subsubsec:optical flare}
\begin{figure}
\centering
\includegraphics[width=0.48\textwidth]{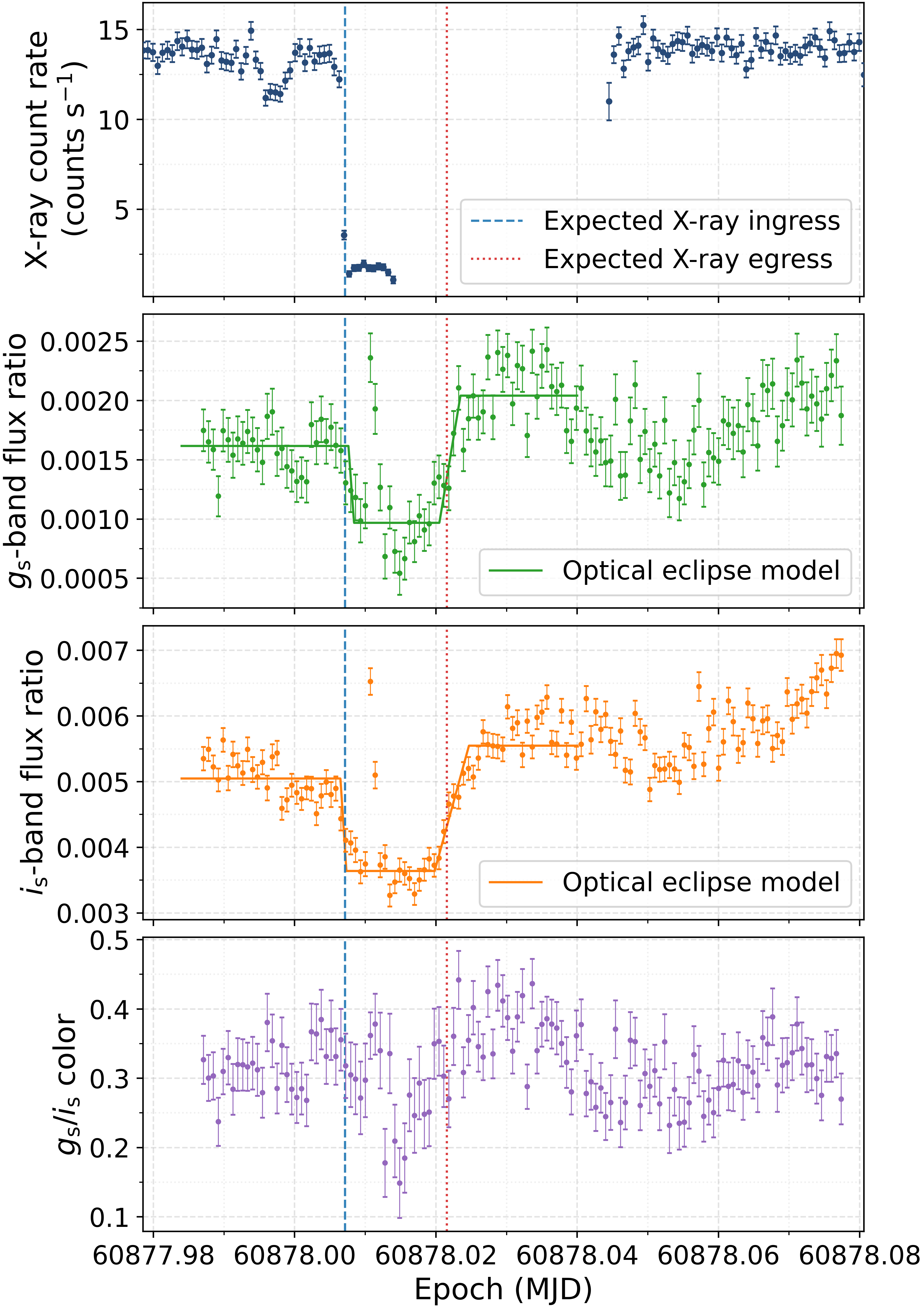}
    \caption{Top panel: X-ray light curve collected by EP/FXT. Second and third panels: ULTRACAM light curves in $g_{\rm s}$-band and $i_{\rm s}$-band, expressed as the flux ratio between the target and the comparison star. The details of the superimposed optical eclipse models (green and orange solid lines) can be found in the text. Bottom panel: the optical color computed as the ratio between $g_{\rm s}$-band and $i_{\rm s}$-band fluxes. In all panels, the expected X-ray ingress (dashed blue lines) and egress (dotted red lines) epochs are calculated from ObsID\,11900304129. The data are rebinned at a time resolution of 60\,s. }
    \label{fig:optical light curves}
\end{figure}

The ULTRACAM light curves reveal an optical flare that occurred entirely during the eclipse and lasted a few minutes (Figure~\ref{fig:optical light curves}). It is likely a real feature and not an artifact as it was visible in both bands, and it does not correlate with variations in the seeing, transparency or object positions on the CCD. The flare reached a peak intensity exceeding the out-of-eclipse level of the persistent optical emission. The color evolution, shown in the bottom panel of Figure~\ref{fig:optical light curves}, indicates that the emission becomes slightly bluer at the flare peak. 

\section{Discussion} \label{sec:discussion}
\src\ is a new NS-LMXB discovered in outburst by the \ep\ on 2025 June 23. It exhibited rare behaviors including clocked bursts, eclipses and dip events. In this section we discuss in detail these phenomena and their implications, as well as the spectral properties of the persistent emission.

\subsection{Properties of the binary system} \label{subsec:binary properties}
With the knowledge of the orbital period determined from the eclipses, i.e. $P_{\rm orb} \approx 6.48$\,hr (see Section~\ref{subsec:eclipse&dip}), we can give a rough estimate of the mass and radius of the companion star. 
\citet{Eggleton1983} derived the effective radius of the Roche lobe ($R_{\rm L}$) as 
\begin{equation}
    \frac{R_{\rm L}}{a} = \frac{0.49q^{2/3}}{0.6q^{2/3}+{\rm ln}(1+q^{1/3})},
\end{equation}
where $q = M_2/M_1$, with $M_1$ and $M_2$ being the masses of the NS and the companion, respectively. $a$ is the orbital separation of the system given by 
\begin{equation}
    \frac{a}{R_\odot} \approx 0.51\left(\frac{M_1+M_2}{M_\odot}\right)^{1/3}P_{\rm orb}^{2/3}.
\end{equation}
Assuming a Roche-lobe filling companion with $R_2\approx R_{\rm L}$, where $R_2$ is the radius of the star, these yield a mass-radius relation for the companion,
\begin{equation}
    \frac{R_{\rm 2}}{R_\odot} = \frac{0.245\beta q^{2/3}((M_1+M_2)/M_\odot)^{1/3}P_{\rm orb}^{2/3}}{0.6q^{2/3}+{\rm ln}(1+q^{1/3})}.
\end{equation}
To solve for both $R_2$ and $M_2$, another mass-radius relation is required, for which we tested three relations available in literature: the empirical mass-radius relation for single low-mass ($M_2 < 1.66 M_\odot$) main-sequence stars \citep{Demircan1991}, the mass-radius relation derived with a sample of companion stars in cataclysmic variables (CVs; but LMXBs appear to follow a similar trend; \citealt{Smith1998}), and a more recent relation which also applies to Roche-lobe filling CV donors \citep{McAllister2019}. Assuming a typical NS mass of $M_1=1.4M_\odot$, the three relations give broadly consistent values of $M_2\approx0.6-0.8M_\odot$ and $R_2\approx0.7-0.8R_\odot$. These are consistent with a K-type star.\footnote{Based on the relation between stellar mass and spectral type provided in \href{http://www.pas.rochester.edu/~emamajek/EEM_dwarf_UBVIJHK_colors_Teff.txt}{A Modern Mean Dwarf Stellar Color and Effective Temperature Sequence}. See also \citet{Pecaut2013}. } Moreover, \citet{Smith1998} provided an empirical relation between the companion spectral type and the orbital period for CVs and LMXBs, which, with $P_{\rm orb} \approx 6.48$\,hr, predicts a spectral type of K7 with a range of K4 to M0, confirming our previous estimate. 

The inclination of an eclipsing system can be expressed by
\begin{equation}
    {\rm sin}i \approx \frac{\sqrt{1-(R_2/a)^2}}{{\rm cos}\theta_{{\rm ec},X}},
\end{equation}
where $\theta_{{\rm ec},X}$ is the eclipse half-angle, which can be estimated as $\theta_{{\rm ec},X} = \pi D_{\star,X}/P_{\rm orb}$ \citep{Joss1984}. From the above, we estimate a high binary inclination of $i\approx73-75^\circ$. 

The longer and shallower eclipses seen in the optical bands indicate that a substantial fraction of the optical emission originates in an extended region surrounding the NS. The characteristic radius of the eclipsed emitting region at wavelength $\lambda$ is given by
\begin{equation}
R_{\lambda} = a\sqrt{\,1 - \sin^2 i \,\cos^2\!\theta_{\rm ec, \lambda}} - R_2.
\end{equation}
We obtain $R_{g_{\rm s}}\simeq(1.39$--$4.87)\times10^9\,{\rm cm}$ and $R_{i_{\rm s}}\simeq(4.17$--$10.44)\times10^9\,{\rm cm}$, indicating a larger characteristic size for the redder optical emission. The eclipse depths are likewise consistent with a progressively larger extended/uneclipsed contribution at longer wavelengths.

The optical flare detected during eclipse (see Section~\ref{subsubsec:optical flare}) also provides additional clues about the companion. The flare properties resemble impulsive magnetic events in active late-type stars, in which a brief heating episode produces a hot, blue continuum followed by a more gradual cooling phase. The fact that the flare remains visible during eclipse implies that its emitting region is not obscured together with the NS and the inner accretion flow. This points to the hemisphere of the Roche-lobe–filling companion facing the observer as the most likely site of emission. Overall, this would be consistent with our estimate of a companion star with a spectral type of K4 to M0, which is known to produce such energetic flares from magnetic activities \citep[e.g.][]{Yang2023}. However, the possibility of the optical flare originating from the reprocessed emission of a type-I X-ray burst cannot be completely ruled out. For example, a recent study by \citet{Rikame2025} showed that type-I X-ray bursts can indeed be detected during eclipses as reprocessed emission, although in their cases the emission can be detected in the X-ray band, while for the optical flare under discussion here, no simultaneous increase in X-rays was detected.

\subsection{Constraints on the distance and luminosity} \label{subsec:distance}
From the broadband spectral analysis with \nustar\ and FXT data (Table~\ref{tab:fit_broadband}), we constrained the value of the hydrogen column density to $N_{\rm H} = 0.48^{+0.08}_{-0.07} \times 10^{22}$\,cm$^{-2}$. This result was then compared with the estimated column density curve along the line of sight in the direction of \src, which is accessible from the 3D-$N_{\rm H}$-tool\footnote{\url{http://astro.uni-tuebingen.de/nh3d/nhtool}} \citep[][]{Doroshenko2024, Edenhofer2024, Yao2017, Planck2016}, to estimate the source distance, $d$. We used a Monte Carlo method to account for the uncertainties in $N_{\rm H}$ and derived a lower limit of $d_{{\rm min},X}\sim 1.6\,{\rm kpc}$ for the distance. The peak bolometric flux of the type-I X-ray bursts we estimated in this work is $\sim2.6\times10^{-9}\,\flux$ (Section~\ref{subsubsec:burst spec}). The Eddington luminosity for the bursts is $L_{\rm Edd,bst}\approx3.18\times10^{38}\,\lum$, estimated with an NS mass of $1.4\msun$ and a hydrogen mass fraction of $\sim0.1$ (See Section~\ref{subsec:comparison with others} for discussion on the burst fuel). By limiting the peak burst luminosity under the Eddington value, and assuming the emission is isotropic, we obtained a loose upper limit of $d_{{\rm max},X}\sim32.0\,{\rm kpc}$ on the distance. This value is not constraining as it already exceeds the approximate extent of the Galaxy in the direction of \src, which is $\sim25\,{\rm kpc}$ \citep[e.g.][]{Castro-Ginard2021}. 

A better constraint on the distance can be derived using the optical data obtained by ULTRACAM. During the optical eclipse (Section~\ref{subsubsec:optical eclipse}), it is reasonable to assume that the optical emission is predominantly from the non-irradiated side of the companion star, and therefore the apparent magnitude of the star can be measured. Using SMSS\,171158.71$-$333340.7, a bright star approximately $1.1\arcmin$ south of the target, as a calibrator, the observed optical fluxes were calibrated into magnitudes. We adopted the faintest magnitudes during the eclipse as the apparent magnitudes of the star, which are $m_{g_{S}}=24.4\pm2.9$ (1$\sigma$ uncertainty) in the $g_{\rm s}$-band and $m_{i_{S}}=20.8\pm0.2$ in the $i_{\rm s}$-band. Then, tentatively assuming a spectral type of K4 to M0 (Section~\ref{subsec:binary properties}), the absolute magnitudes of the companion in the two bands were derived to be $M_{g_{S}}\approx7.5-9.6$ and $M_{i_{S}}\approx6.3-7.6$.\footnote{The absolute magnitude in the $V$-band is provided in \href{http://www.pas.rochester.edu/~emamajek/EEM_dwarf_UBVIJHK_colors_Teff.txt}{A Modern Mean Dwarf Stellar Color and Effective Temperature Sequence} and the value can be transformed into those in SDSS photometry \citep[see][]{jordi2006}. Note that ULTRACAM uses ``Super SDSS filters'', but here we tentatively ignored the difference for simplicity. } Further considering the interstellar extinction in the two bands, $A_{g_{S}}(d)$ and $A_{i_{S}}(d)$,\footnote{The extinction curve in the $V$-band was obtained with the \href{http://astro.uni-tuebingen.de/nh3d/nhtool}{3D-$N_{\rm H}$-tool}, which was then transformed into $A_{g_{S}}(d)$ and $A_{i_{S}}(d)$ using \texttt{python} package \href{https://dust-extinction.readthedocs.io/en/latest/}{\texttt{dust\_extinction}} \citep{Gordon2024,Gordon2023,Gordon2009,Fitzpatrick2019,Gordon2021,Decleir2022}. } the distance of the system can be estimated by solving the following equation:
\begin{equation}
    m_{\lambda} - M_{\lambda} = 5\,{\rm log}_{10}(d/10{\rm pc}) + A_{\lambda}(d). 
\end{equation}
We thus derived $d_{g_{\rm s}}>1.4\,{\rm kpc}$ and $d_{i_{\rm s}}\sim2.0-6.4\,{\rm kpc}$ from the $g_{\rm s}$- and $i_{\rm s}$-band data, respectively. The latter provides a tighter constraint, as the data have smaller uncertainties. 

In summary, the best constraints on the distance of the system available so far are $d_{\rm min}\sim2.0\,{\rm kpc}$ and $d_{\rm max}\sim6.4\,{\rm kpc}$, with which the luminosity of the source can be estimated. Extrapolating the best-fit 1--10\,keV spectral model presented in Table~\ref{tab:fit_broadband} to the 0.01--100\,keV range, we estimated a bolometric correction factor of $K_{\rm bol}\sim3$. The bolometric flux of the persistent emission at the epoch of the \nustar\ observation is therefore $F_{\rm bol}\sim F_{0.01-100{\rm keV}}\sim6\times10^{-10}\,\flux$, which corresponds to a luminosity of $L_{\rm bol}\sim2.8\times10^{35}(d/2.0\,{\rm kpc})^2\,\lum=2.9\times10^{36}(d/6.4\,{\rm kpc})^2\,\lum$. The Eddington ratio is thus $L_{\rm bol}/L_{\rm Edd,acc}\sim0.1\%-1.5\%$, where $L_{\rm Edd,acc}\approx2.0\times10^{38}\,\lum$ is the Eddington luminosity estimated with an NS mass of $1.4\msun$ and a hydrogen mass fraction of $\sim0.75$ assumed for the accretion flow from a hydrogen-rich K-type companion (Section~\ref{subsec:binary properties}). 

We note that this estimated value of the Eddington ratio is significantly lower than that of a typical LMXB in outburst, which is of the order of a few tens of percent \citep[e.g.][]{Munoz-Darias2014}. The burst recurrence time of $\sim8200\,{\rm s}$ also points to a higher bolometric luminosity of $\lesssim10\%L_{\rm Edd}$ (Section~\ref{subsec:comparison with others}). On the one hand, it could imply that we have underestimated the source distance. One possibility could be that the optical emission during the eclipse is still dominantly contributed by the disc, and the true apparent magnitude of the companion star is therefore higher than the value we adopted, which then would lead to a larger distance estimate. On the other hand, it is also possible that the low observed luminosity is a result of the high inclination of \src. For example, 4U\,1822$-$371, another high-inclination LMXB with a low observed luminosity, was argued to have an intrinsic luminosity which is two orders of magnitude higher, and the suggested explanation is that most of the observed radiation is contributed by the photons scattered along the line of sight in an accretion-disc corona, which is only a small fraction of those emitted in the inner accretion flow \citep[see e.g.][and references therein]{Anitra2021}. Moreover, high inclination has been invoked in the past to explain the behavior of LMXBs which display faint (i.e., peaking at $\sim1\%$ Eddington ratio or less) outbursts, the so-called very faint X-ray transients \citep[e.g.][]{Muno2005, King2006}.

\subsection{The dip events} \label{subsec:dips}
We identified a total of six dip events in our main dataset (Section~\ref{subsec:eclipse&dip}). Despite the significant differences in the duration and the short-term variations seen in the light curves, we note that all six dips were found to occur within the orbital phase range of $\sim$0.6--0.7 (with the center of eclipses defined as phase 0.5), suggesting a possible correlation with the binary orbit. The phase dependency has also been seen in other dipping systems \citep[for a review, see][]{DiSalvo2023}. This supports the interpretation that dip events arise from the obscuration of the central X-ray source by the bulge produced by the impact of matter transferred from the companion star at the outer disk. Alternatively, it is also possible that the obscuration is not resulted from the bulge itself, but it is instead caused by the mass transfer stream from the star, which could ricochet off the disk edge during the impact and overflow towards smaller radii, generating the observed dips \citep{Armitage1996}. This scenario was used to explain the phase-dependent dips observed in GRO\,J1655$-$40 \citep{Kuulkers2000}, although the dips in this system happened at a different phase range from that of \src\ (note that \citealt{Kuulkers2000} defined the mid-eclipse as phase 1.0). However, the occurrence of dips is still far from regular, e.g., the \nustar\ observation covered the 0.6--0.7 range of orbital phase also during two other orbits, yet no dips were detected, suggesting a variable nature of the obscuring material. 

We also note that the dips were observed only in the observations before 2025 June 28 and in the last observation on 2025 July 13, which coincide with the periods when the source flux was at a relatively low value of $1.3\times10^{-10}\,\flux$ (Figure~\ref{fig:fxt_spec_evolution}). The flux dependence of dip events was also seen in other sources, e.g. MAXI\,J1820$+$070 \citep{Kajava2019} and MAXI\,J1659$-$152 \citep{Kuulkers2013}. The disappearance of dips at higher flux levels likely reflects changes in the ionization and vertical structure of the disk rim: as the accretion rate increases, stronger irradiation and radiation pressure can ionize or geometrically smooth the bulge at the stream–disk impact point, reducing its ability to obscure the central source or to deflect the mass transfer stream. 

Apart from the six dips in the main dataset, a possible optical dip can be seen also in the ULTRACAM light curves (Figure~\ref{fig:optical light curves}) at around MJD\,60878.00 (for $g_{\rm s}$-band), together with an X-ray counterpart in the FXT light curve. Optical dips were also observed in other dipping systems \citep[e.g.][]{Motch1987,Thomas1993,Panizo-Espinar2024}. We note that this dip showed a wavelength-dependent delay, with the detection in the less energetic band lagging behind that in the more energetic band. Moreover, this event happened before the eclipse and therefore at a different orbital phase from the other six dips. The decrease of the X-ray flux during this dip is also not as significant as the others, suggesting a different physical origin of this event.

\subsection{Spectral properties of the persistent emission} \label{subsec:spec discussion}
We extensively monitored the spectral evolution of \src\ with EP/FXT during the first 21 days of the outburst (see Figure~\ref{fig:fxt_spec_evolution}). The source was initially detected with an observed persistent flux of $\sim 1.3\times10^{-10}\,\flux$ in 0.5--10\,keV band, which rose to $\sim 1.8\times10^{-10}\,\flux$ on June 27. The flux remained roughly constant until July 7 then slowly declined back to the initial flux level of $\sim 1.3\times10^{-10}\,\flux$. Despite the variation in flux, the spectral shape did not show significant changes (see also Figure~\ref{fig:all_fxt_spec}). In all the observations presented in this work, \src\ exhibited a typical hard-state spectrum, with a low power-law photon index of $\Gamma\sim 1.5$. Moreover, from our broadband spectral analysis performed with both EP/FXT and \nustar\ data (see Figure~\ref{fig:broadband-spec} and Table~\ref{tab:fit_broadband}), we obtained an inner disk radius of $R_{\rm in}=38^{+14}_{-9}R_{\rm g}$ and a corona electron temperature of $kT_{\rm e}=22^{+20}_{-3}$\,keV. These are also broadly consistent with those seen in the hard spectral state of NS-LMXBs \citep[e.g.][]{Pintore2018, Marino2022, Ludlam2024, Illiano2024}. We thus suggest the following accretion geometry: a truncated accretion disk with the inner region replaced by a hot accretion flow (corona), which Compton scatters soft photons emitted from a hot region on the NS surface. The area of this region is $\sim 4-36\,{\rm km}^2$ as measured from the blackbody normalization, using the distance of the system at $\sim2.0-6.4$\,kpc (Section~\ref{subsec:distance}).

\subsection{Comparison of \src\ with other clocked bursters} \label{subsec:comparison with others}
The defining observational characteristic of clocked bursters is the quasi-periodic recurrence of type-I X-ray bursts. Owing to long data gaps and intrinsic variability in the burst rate, we could not obtain a coherent periodic solution across the full 16-day dataset, which included 16 bursts. Reliable recurrence times were measured only for eight consecutive pairs of bursts spanning 1.6\,days around the time of the \nustar\ observation. Despite the presence of bad time intervals in the data, their durations are much shorter than the characteristic timescale of the bursting cycle. For this subset, we found $t_{\rm rec}=8196 \pm 177\,$s ($\nu_{\rm rec} \approx 0.122$\,mHz). The possibility that multiple bursts were missed during these gaps (potentially leading to an overestimation of $t_{\rm rec}$ by identifying it as a lower-frequency harmonic of a shorter underlying fundamental period) is ruled out by the absence of bursts during intervals with continuous data coverage.
The data showed a marginal linear decrease in the recurrence rate of $\dot{t}_{\rm rec} = -270 \pm 110$\,s\,day$^{-1}$, although notable scatter remains, indicating intrinsic variability in the accretion rate or burning conditions on the NS on the timescale of individual bursts (Figure \ref{fig:trec}). Additional continuous observations covering multiple consecutive bursts would be required to provide a more precise characterization of the source’s temporal behavior.

In the seven known clocked bursters, the recurrence time of bursts is generally found to be anti-correlated with the source flux and is typically of the order of hours (see \citealt{Cavecchi2025} and references therein for a further discussion). This anti-correlation is also found in a larger sample of NS-LMXBs with type-I X-ray bursts, as can be seen in the trend of the distribution of such systems in the $L_{\rm bol}-t_{\rm rec}$ plane, extending from $t_{\rm rec}\sim10^8$\,s at $L_{\rm bol}\sim0.1\%L_{\rm Edd}$ to $t_{\rm rec}\sim10^3$\,s at $L_{\rm bol}\sim100\%L_{\rm Edd}$ \citep[see figure 1 in][]{Kormpakis2025}. Such a trend is expected, since a higher $L_{\rm bol}$ implies a higher accretion rate, reducing the time needed to accumulate enough amount of matter to trigger thermonuclear runaway. Based on this observed trend, $t_{\rm rec}\sim8200$\,s as measured for \src\ would imply a luminosity of $L_{\rm bol}\lesssim10\%L_{\rm Edd}$, which is higher by 1--2 orders of magnitude than the luminosity we measured in Section~\ref{subsec:distance}. We also found an increasing trend for the persistent emission count rate during the \nustar\ observation, which is qualitatively consistent with the decreasing trend of $t_{\rm rec}$ (Section~\ref{subsubsec:burst recurrence}). We further note that the increase in the count rate was energy-dependent, i.e., the count rate in the 3--10\,keV band rose faster than that in the 10--79\,keV band, demonstrating a slow evolution of the spectral shape in real time (Figure~\ref{fig:trec}). 

The spectral properties of the persistent emission are typical of a hard spectral state (Section~\ref{subsec:spec discussion}), which is also the case for several other clocked bursters, including the prototype, GS\,1826$-$24 \citep[e.g.][]{Ubertini1999}. 

We also estimated the $\alpha$ parameter for bursts No. 5, 6 and 10, which has a high value of 120--130 (Section~\ref{subsubsec: alpha}). As a comparison, GS\,1826$-$24 had an $\alpha$ value of $\sim60$ \citep{Ubertini1999}. Further assuming that the mass and radius of the NS are $1.4M_\odot$ and $11.2$\,km, we estimated the mean hydrogen mass fraction at ignition to be $\bar{X}\approx0.09-0.11$ \citep[using equation (11) in][]{Galloway2022}, corresponding to helium-rich burst fuel. For simplicity, here we assumed that the burst and the persistent emission have the same anisotropy factors. Since we expect the accreted gas from the K-type companion to be hydrogen-rich, this $\bar{X}$ value implies that most of the accreted hydrogen is consumed by stable burning before a helium burst ignites \citep[burning regime (III) in][]{Galloway2021}.

\section{Summary} \label{sec:summary}

We summarize the main results on this new transient NS binary system, \src, as derived from X-ray and optical observations:
\begin{itemize}
    \item \src\ is a clocked burster and its burst recurrence time can be characterized over a subset of nine bursts spanning 1.6\,days around the \nustar\ observation, which is $t_{\rm rec}=8196 \pm 177\,$s. A likely linear decrease of its recurrence time ($\dot{t}_{\rm rec} = -270 \pm 110$\,s\,day$^{-1}$) is also detected, although notable scatter remains. Remarkably, the \nustar\ light curve demonstrated an increasing trend, which is qualitatively consistent with the decrease of $t_{\rm rec}$.  
    
    \item \src\ is an eclipsing binary with an orbital period of $P_{\rm orb}=6.48301 \pm 0.00003$\,hr and a typical X-ray eclipse duration of $D_{\star,X} = 1245.5^{+6.9}_{-6.5}$\,s. Apart from the eclipses, the source also exhibits dip events that seem to occur mostly at the same orbital phase range. The presence of the dips also exhibits a possible correlation with the source flux. The mass and radius of the companion star are estimated to be $M_2 \approx 0.6-0.8M_\odot$ and $R_2 \approx 0.7-0.8 R_\odot$, pointing to a K-type star. The binary inclination is $i\approx73-75^\circ$.

    \item Eclipses and dips are also present in the ULTRACAM optical light curves. The optical eclipses are broader and shallower than in X-rays and show a clear wavelength dependence, with longer and less deep eclipses in the redder band. By comparing the optical and X-ray ingress and egress times, we infer that a substantial fraction of the optical emission arises from an extended region in the accretion flow (disk and/or stream impact region). The dip observed in this joint EP-ULTRACAM observation also showed a wavelength-dependent delay, with the detection in the less energetic band lagging behind that in the more energetic band. In addition, this dip probably has a different physical origin from the other dips, as it occurred in a different orbital phase and the X-ray flux decrease during it is also less significant. 
    
    \item The ULTRACAM data reveal a short optical flare occurring during the eclipse. The event shows a blueward color change at peak and lacks an X-ray counterpart, pointing to a magnetic flare on the Roche-lobe-filling companion star, even though the possibility of the optical flare originating from the reprocessed emission of a type-I X-ray burst cannot be fully ruled out. 
    
    \item The distance of the system is constrained to $\sim 2.0-6.4$\,kpc. With this distance range, the Eddington luminosity ratio of the persistent emission of \src\ during the \nustar\ observation is estimated to be $L_{\rm bol}/L_{\rm Edd,acc}\sim0.1\%-1.5\%$. Such a low value of observed luminosity could be the result of the high inclination of the system. 
    
    \item During the first 21 days of the outburst, the persistent emission of \src\ showed moderate variation in the observed flux, but the spectral shape did not show significant changes. The spectral properties are typical of a hard spectral state. 
    
    \item The ratio of accretion energy to thermonuclear energy ($\alpha$) is estimated to be 120--130, which corresponds to a mean hydrogen mass fraction at ignition of $\bar{X}\approx0.09-0.11$, implying helium bursts with the accreted hydrogen being depleted by stable burning in-between bursts. 
\end{itemize}

\begin{acknowledgments}
This work is based on the data obtained with the \ep, a space mission led by the Chinese Academy of Sciences, in collaboration with the European Space Agency, the Max Planck Institute for Extraterrestrial Physics (Germany), and the Centre National d'Études Spatiales (France). This work is supported by the National Natural Science Foundation of China (Grant Nos. 12333004 and 12433005), and the Strategic Priority Research Program of the Chinese Academy of Sciences (Grant No. XDB0550200). This work is also based on the data obtained with: the \nustar\ mission, a project led by the California Institute of Technology, managed by the Jet Propulsion Laboratory, and funded by NASA; and ULTRACAM on the 3.5-m NTT at La Silla, Chile. 

YLW is supported by the China Scholarship Council (No. 202404910397). AM and NR are supported by the European Research Council (ERC) under the European Union’s Horizon Europe research and innovation programme (ERC Consolidator Grant ``MAGNESIA'' No. 817661, and ERC Proof of Concept ``DeepSpacePULSE'' No. 101189496; PI: NR). FCZ is supported by a Ram\'on y Cajal fellowship (grant agreement RYC2021-030888-I). EP is supported by a Juan de la Cierva fellowship (JDC2022-049957-I). YLW, FCZ, EP, AM and NR are also supported by the Catalan grant SGR2021-01269, Spanish grant PID2023-153099NA-I00 (PI: FCZ) and Unidad de Excelencia Maria de Maeztu CEX2020-001058-M. VSD and ULTRACAM are supported by STFC grant ST/Z000033/1. JB-P and IR acknowledge financial support from Spanish grants PID2021-125627OB-C31 and PID2024-158486OB-C31 funded by MCIU/AEI/10.13039/501100011033 and by ``ERDF A way of making Europe'', by the Generalitat de Catalunya/CERCA programme, and by the European Research Council (ERC) under the European Union’s Horizon Europe programme (ERC Advanced Grant SPOTLESS; No. 101140786). JB-P acknowledges support from Spanish grant PRE2022-101942. AP acknowledges support from grants PID2024-155316NB-I00, PID2021-124581OB-I00, and 2021SGR00426. YC acknowledges support from the grant RYC2021-032718-I, financed by MCIN/AEI/10.13039/501100011033 and the European Union NextGenerationEU/PRTR and funds from the Spanish MINECO (PID2023-148661NB-I00)/E.U. FEDER (PI Jos\'e). SG acknowledges the support of the CNES. This work also received financial support from INAF through the GRAWITA 2022 Large Program Grant. YX acknowledges support from National Science Foundation of China through grant NSFC-12521005. 
\end{acknowledgments}

\facilities{\ep\ (WXT and FXT), \nustar, ESO-NTT (ULTRACAM). }

\software{FXTDAS v1.20 \citep{Zhao2025}, HEASOFT v6.35.2 \citep{heasoft}, Matplotlib v3.10 \citep{hunter07}, PINT \citep{Luo2021}, XSPEC v12.15.0 \citep{arnaud96}, NuSTARDAS v.2.1.4a. }

\appendix

\section{Measurement of the burst recurrence time}\label{appendix:t_rec}
To characterize $t_{\rm rec}$, we define burst times of arrival (TOAs) corresponding to the burst start times. TOA uncertainties were estimated from the flux rise duration, i.e. from the start time to the peak time. 

First, we estimated $t_{\rm rec}$ using the seven detected bursts in the \nustar\ observation (see Figure~\ref{fig:nustar_lc}) as this segment of the dataset exhibited the highest burst detection rate. Assuming a stable and periodic $t_{\rm rec}$ across the \nustar\ observation, we estimated the underlying periodicity by computing all pairwise TOA differences and identifying a common integer divisor. This yielded a period of approximately 8200\,s, corresponding to a burst recurrence frequency $\nu_{\rm rec}$ of $\approx0.122$\,mHz. 

Next, we used the estimated $\nu_{\rm rec}$ to attempt to derive a coherent timing solution across all 16 burst TOAs using the \texttt{PINT} pulsar timing software \citep{Luo2021,pint24}. We first fitted the data with a model in which $\nu_{\rm rec}$ was the sole free parameter. We were unable to derive an acceptable solution --- post-fit residuals spanned thousands of seconds, indicating the presence of intrinsic variation in the burst repetition frequency. We also found degeneracies in the inferred number of burst cycles between consecutive detections when the temporal gap exceeds approximately ten cycles. Introducing a first frequency derivative, $\dot{\nu}_{\rm rec}$, as an additional free parameter did not produce a statistically significant improvement in the fit. 

Lacking a coherent solution across all bursts, we opted to conservatively assess the timing variability of bursts by applying a timing model with $\nu_{\rm rec}$ as the sole free parameter to pairs of consecutive bursts across the full TOA set. Measurements from TOA pairs yielding degenerate solutions were excluded from further analysis. Solutions for 8 of the 15 TOA pairs, covering all \nustar\ bursts, were obtained over a 1.6-day interval, from the EP/FXT burst on 2025-06-27 19:23:31 (UTC) to that on 2025-06-29 09:53:52 (UTC), corresponding to bursts No.~3--11 in Table~\ref{tab:bursts}. Best-fit $t_{\rm rec}$ values are shown in the left panel of Figure~\ref{fig:trec}. 

From these eight $t_{\rm rec}$ measurements we obtained a weighted median $t_{\rm rec}=8196 \pm 177\,$s (95\% confidence level). A simple weighted constant fit gave post-fit residuals with reduced chi-squared $\chi^2_{\rm red} = 8.1$ (7 dof), indicating unmodeled variability in the burst recurrence time. A decreasing trend in $t_{\rm rec}$ can be seen, and fitting the temporal evolution with a linear model yielded $\dot{t}_{\rm rec} = -270 \pm 110$\,s\,day$^{-1}$ (95\% confidence level), and a corresponding $\chi^2_{\rm red} = 3.2$ (6 dof). To account for the extra scatter beyond the quoted statistical uncertainties, we modeled the $t_{\rm rec}$ values as Gaussian-distributed around either a constant or a linear trend, with an additional intrinsic scatter term $\sigma_{\rm int}$ added in quadrature to the individual statistical errors, and fitted both models by maximum likelihood estimation. Under the constant model we obtained $\mu = 8212\rm\,s$ and $\sigma_{\rm int} \simeq 130\rm\,$s, while the linear model gave $\mu = 8338\rm\,s$, $\dot{t}_{\rm rec} = -234\rm\,s\,day^{-1}$ and $\sigma_{\rm int} \simeq 70$\,s. A likelihood-ratio test between these two models, calibrated with a parametric bootstrap using 2000 simulations, yielded a $p$-value of $\simeq 0.02$ for a non-zero slope, and a bootstrap 95\% confidence interval for the slope of $[-3.6,-1.1]\times10^{2}$\,s\,day$^{-1}$, indicating modest but statistically significant evidence for a decrease of $t_{\rm rec}$ during the outburst.

\section{Supplementary results} \label{appendix:supplements}
In this appendix, we present supplementary tables and figures to complement the results reported in the main text. The details of the X-ray observations presented in this work are listed in Table~\ref{tab:log}. The properties of the 16 detected type-I X-ray bursts are listed in Table~\ref{tab:bursts}. The fitting results of the relation between $t_{\rm rec}$ and the observed \nustar\ count rate during persistent emission are presented in Figure~\ref{fig:trec_flux_fit}. The light curves and spectra collected by EP/FXT are respectively presented in Figure~\ref{fig:all_fxt_lc} and Figure~\ref{fig:all_fxt_spec}. The results of the time-resolved spectral analyses for the bursts detected by \nustar\ and EP/FXT are shown in Figure~\ref{fig:NuSTAR_burst_time_res} and Figure~\ref{fig:FXT_burst_time_res}, respectively. Burst No.~8 is the only burst simultaneously covered by both EP/FXT and \nustar, and its spectral analysis results are presented separately in Figure~\ref{fig:burst8}. 

\renewcommand{\thefigure}{B\arabic{figure}}
\renewcommand{\thetable}{B\arabic{table}}
\renewcommand{\theHfigure}{B\arabic{figure}}
\renewcommand{\theHtable}{B\arabic{table}}
\setcounter{figure}{0}
\setcounter{table}{0}

\begin{deluxetable*}{lcccc}
\tablewidth{0pt}
\tablecaption{Journal of the X-ray observations of \src\ presented in this work \label{tab:log}}
\tablehead{
\colhead{Telescope} & \colhead{Mode} & \colhead{ObsID} & \colhead{Start -- End Time (UTC)} & \colhead{Exposure} \\
\colhead{} & \colhead{} & \colhead{} & \colhead{(YYYY-MM-DD HH:MM:SS)} & \colhead{(ks)}
}
\startdata
EP/FXT & FF+FF  & 06800000695 & 2025-06-25 14:09:08 -- 2025-06-25 15:00:09 & 3.1 \\
EP/FXT & PW+TM  & 06800000697 & 2025-06-26 14:08:34 -- 2025-06-26 16:35:19 & 6.0 \\
EP/FXT & PW+TM  & 06800000702 & 2025-06-27 06:08:10 -- 2025-06-27 06:18:35 & 0.6 \\
EP/FXT & PW+TM  & 06800000698 & 2025-06-27 10:56:04 -- 2025-06-27 13:22:38 & 6.1 \\
EP/FXT & FF+FF  & 06800000705 & 2025-06-27 18:55:52 -- 2025-06-27 21:22:22 & 4.8 \\
\nustar &  --   & 81101345002 & 2025-06-27 20:50:57 -- 2025-06-28 21:14:06 & 46.7 \\
EP/FXT & FF+FF  & 06800000706 & 2025-06-28 01:19:42 -- 2025-06-28 03:46:09 & 3.1 \\
EP/FXT & PW+TM  & 06800000699 & 2025-06-28 10:55:28 -- 2025-06-28 13:22:06 & 6.1 \\
EP/FXT & FF+FF  & 06800000707 & 2025-06-28 17:19:19 -- 2025-06-28 18:09:41 & 3.0 \\
EP/FXT & FF+FF  & 06800000708 & 2025-06-29 01:19:07 -- 2025-06-29 03:45:22 & 3.3 \\
EP/FXT & PW+TM  & 06800000700 & 2025-06-29 09:18:55 -- 2025-06-29 11:45:22 & 6.1 \\
EP/FXT & PW+TM  & 06800000701 & 2025-06-30 17:18:07 -- 2025-06-30 19:44:17 & 5.1 \\
EP/FXT & PW+TM  & 06800000717 & 2025-07-01 15:41:30 -- 2025-07-01 18:07:31 & 6.0 \\
EP/FXT & PW+TM  & 06800000718 & 2025-07-02 06:05:06 -- 2025-07-02 08:31:01 & 5.2 \\
EP/FXT & PW+TM  & 06800000720 & 2025-07-03 14:04:08 -- 2025-07-03 16:29:54 & 6.0 \\
EP/FXT & PW+TM  & 06800000722 & 2025-07-05 09:14:35 -- 2025-07-05 11:40:20 & 6.0 \\
EP/FXT & PW+TM  & 06800000723 & 2025-07-06 10:50:04 -- 2025-07-06 13:15:25 & 6.0 \\
EP/FXT & PW+TM  & 06800000724 & 2025-07-07 12:24:40 -- 2025-07-07 14:50:28 & 6.0 \\
EP/FXT & PW+TM  & 11900302977 & 2025-07-09 05:59:06 -- 2025-07-09 08:24:56 & 6.0 \\
EP/FXT & PW+TM  & 11900304129 & 2025-07-10 08:16:45 -- 2025-07-10 11:25:05 & 5.8 \\
EP/FXT & PW+TM  & 11900305153 & 2025-07-11 11:03:37 -- 2025-07-11 13:11:02 & 4.9 \\
EP/FXT & PW+TM  & 11900306945 & 2025-07-12 13:01:38 -- 2025-07-12 16:13:15 & 5.6 \\
EP/FXT & PW+TM  & 11900307969 & 2025-07-13 15:49:35 -- 2025-07-13 17:57:19 & 4.3 \\
\enddata
\tablecomments{FF: Full Frame mode; PW: Partial Window mode; TM: Timing mode. PW+TM denotes the configuration in which FXT-A and FXT-B are set in PW and TM, respectively. }
\end{deluxetable*}

\begin{deluxetable*}{cccccccccc}
\tablewidth{0pt}
\setlength{\tabcolsep}{1.5pt}
\tablecaption{
Properties of the detected bursts. The date (UTC) corresponds to the burst times of arrival (TOAs; see Appendix~\ref{appendix:t_rec}). Decay timescales were derived by fitting each burst with a Gaussian-rise plus exponential-decay model. The fluence is the total number of photons detected within the burst duration. The last column, $t_{\rm rec}$, is the burst recurrence time (see Section~\ref{subsubsec:burst recurrence}), defined as the time elapsed between the start of the burst and the start of the previous burst. Associated 1$\sigma$ uncertainties are indicated in parentheses on the last digit. 
\label{tab:bursts}}
\tablehead{\colhead{Burst No.} & \colhead{Telescope} & \colhead{ObsID} & \colhead{Start Date} & \colhead{Start Time} & \colhead{Peak Time} & \colhead{Decay Timescale} & \colhead{Fluence} & \colhead{$t_{\rm rec}$}\\ \colhead{ } & \colhead{ } & \colhead{ } & \colhead{(YYYY-MM-DD)} & \colhead{(HH:MM:SS)} & \colhead{(HH:MM:SS)} & \colhead{(s)} & \colhead{(counts)} & \colhead{(s)}} 
\startdata
1 & EP/FXT & 06800000695 & 2025-06-25 & 14:30:41 & 14:30:58 & 21(3) & 988(3) & -- \\
2 & EP/FXT & 06800000697 & 2025-06-26 & 14:25:40 & 14:25:57 & 27(3) & 959(5) & -- \\
3 & EP/FXT & 06800000705 & 2025-06-27 & 19:23:31 & 19:23:42 & 40(4) & 796(5) & -- \\
4 & \nustar\ & 81101345002 & 2025-06-27 & 21:43:39 & 21:43:48 & 17(3) & 826(6) & 8408(47) \\
5 & \nustar\ & 81101345002 & 2025-06-28 & 00:00:35 & 00:00:40 & 19(2) & 1549(7) & 8216(41) \\
6 & \nustar\ & 81101345002 & 2025-06-28 & 02:19:45 & 02:19:58 & 42(3) & 1724(3) & 8350(59) \\
7 & \nustar\ & 81101345002 & 2025-06-28 & 06:59:13 & 06:59:30 & 29(3) & 1688(8) & 8384(60) \\
8 & \nustar\ & 81101345002 & 2025-06-28 & 11:31:30 & 11:31:47 & 31(3) & 1939(4) & 8168(59) \\
8 & EP/FXT & 06800000699 & 2025-06-28 & 11:31:46 & 11:31:57 & 30(3) & 1190(6) & 8168(59) \\
9 & \nustar\ & 81101345002  & 2025-06-28 & 15:59:12 & 15:59:23 & 30(3) & 1824(11) & 8031(59) \\
10 & \nustar\ & 81101345002 & 2025-06-28 & 18:13:57 & 18:14:08 & 21(3) & 1739(8) & 8085(54) \\
11 & EP/FXT & 06800000700 & 2025-06-29 & 09:53:52 & 09:53:59 & 32(3) & 1255(10) & 8056(51) \\
12 & EP/FXT & 06800000718 & 2025-07-02 & 06:25:32 & 06:25:45 & 30(3) & 1559(8) & -- \\
13 & EP/FXT & 06800000720 & 2025-07-03 & 16:11:59 & 16:12:14 & 52(4) & 2066(9) & -- \\
14 & EP/FXT & 06800000724 & 2025-07-07 & 14:00:46 & 14:00:55 & 47(3) & 2227(12) & -- \\
15 & EP/FXT & 11900304129 & 2025-07-10 & 10:58:49 & 10:59:14 & 35(3) & 2043(6) & -- \\
16 & EP/FXT & 11900305153 & 2025-07-11 & 11:07:44 & 11:08:01 & 28(3) & 2313(7) & -- \\
\enddata
\end{deluxetable*}
\begin{figure}
\centering
\includegraphics[width=0.48\textwidth]{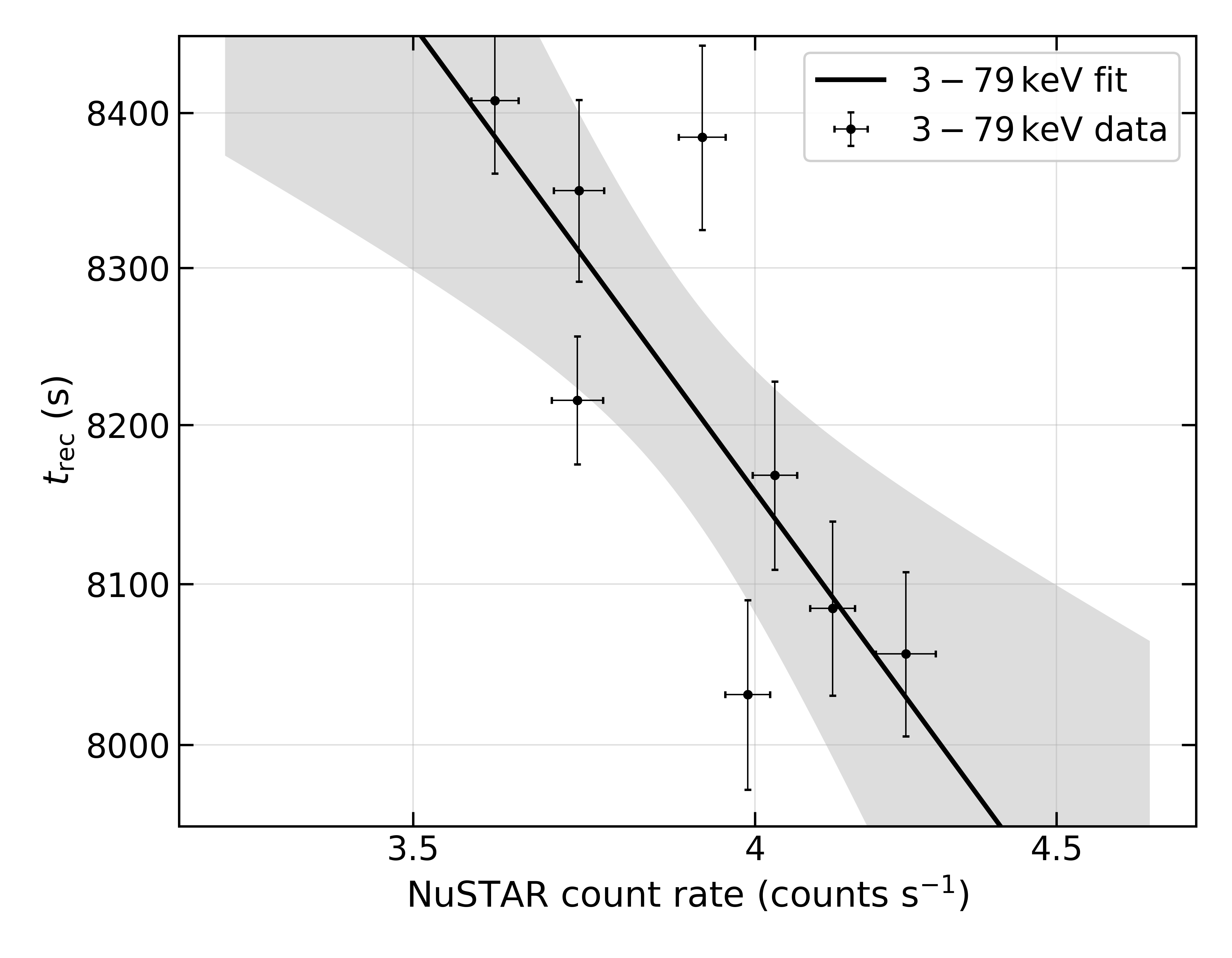}
    \caption{Fitting results of the relation between $t_{\rm rec}$ and the persistent emission count rate for the eight consecutive pairs of bursts with unambiguous cycle counts around the \nustar\ observation (Section~\ref{subsubsec:burst recurrence}). Black circles and solid line represent the \nustar\ data and the corresponding power-law fit ($t_{\rm rec} \propto F^{-0.27\pm0.16}$, $\chi^2_{\rm red} = 3.0$ with 6 dof) in the 3--79\,keV band. Shaded area denotes uncertainties at 95\% confidence level. }
    \label{fig:trec_flux_fit}
\end{figure}
\begin{figure*}
\centering
\includegraphics[width=\textwidth]{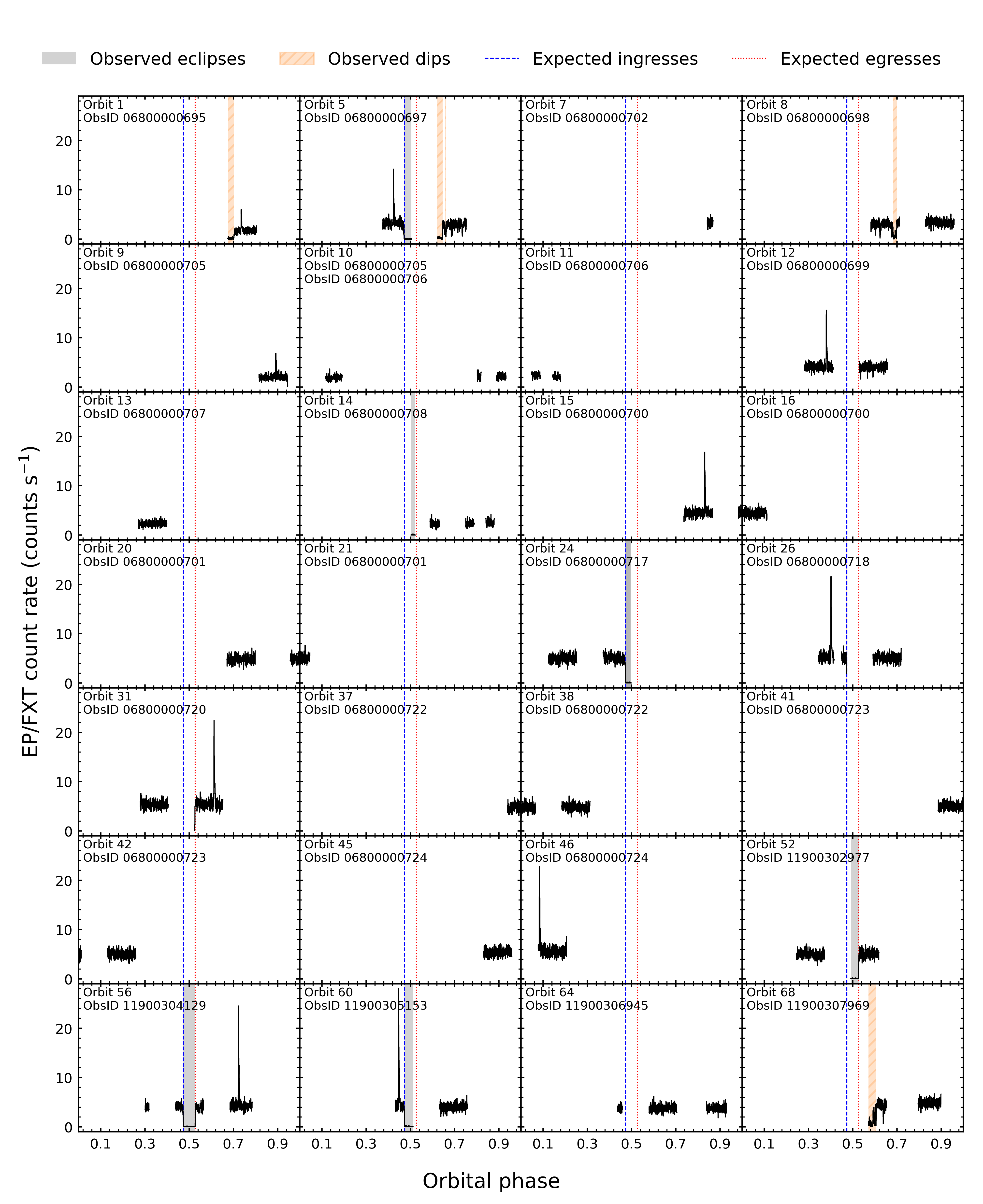}
    \caption{Phase-resolved light curves of all 22 EP/FXT observations presented in this work. Each panel shows the data obtained in one orbit of the binary system. The orbit observed by the first FXT observation is defined as Orbit 1. The expected ingress (dashed blue lines) and egress (dotted red lines) epochs are calculated from ObsID\,11900304129. The shaded gray areas are the observed eclipse periods. The orange hatched bands are the observed dips. Only the data collected by FXT-A are shown.}
    \label{fig:all_fxt_lc}
\end{figure*}
\begin{figure}
\centering
\includegraphics[width=0.48\textwidth]{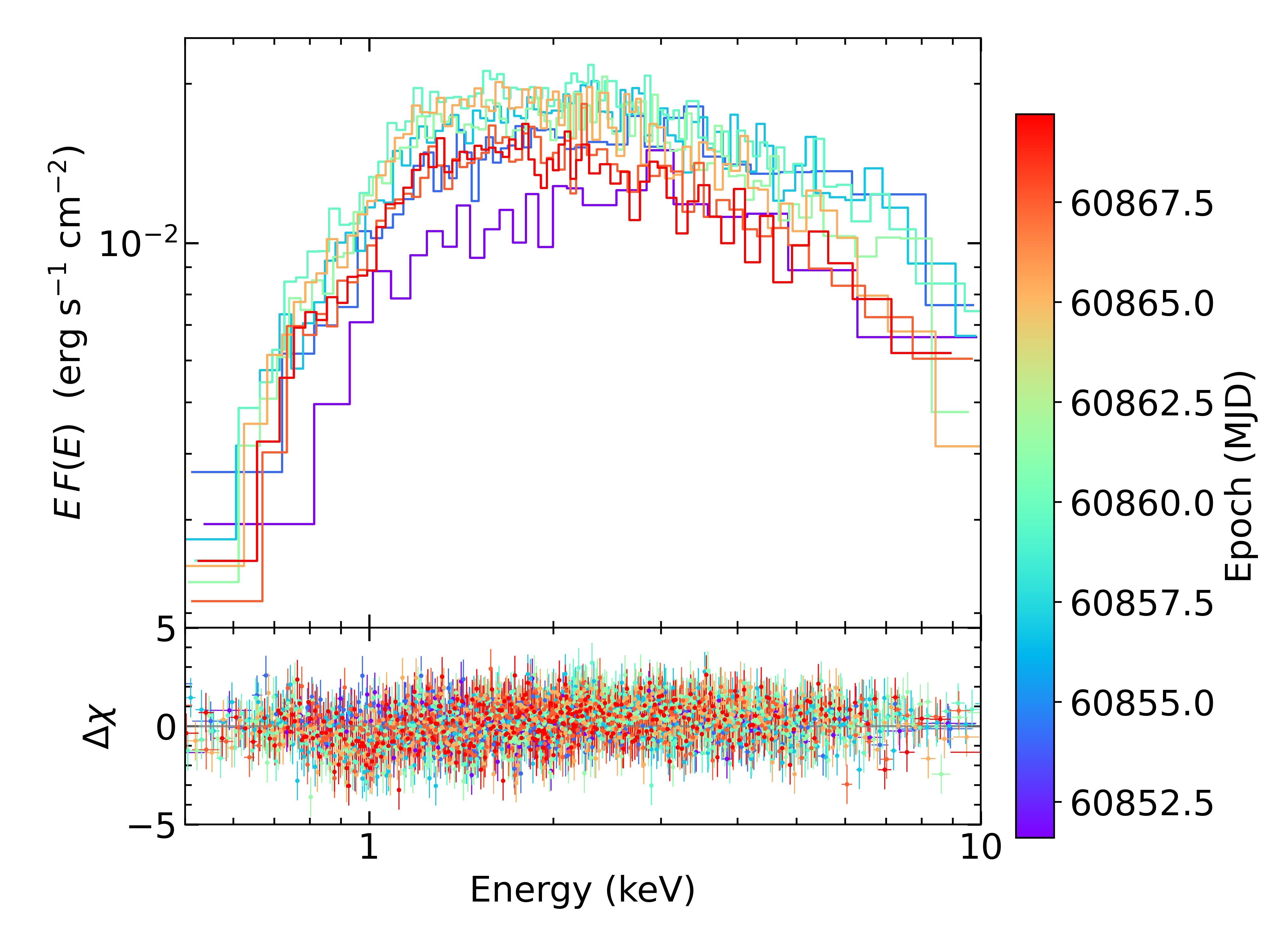}
    \caption{Evolution of the unfolded spectra of \src\ as observed by EP/FXT (upper panel) and the spectral fitting residuals (lower panel). The model adopted for spectral analysis is $\texttt{TBabs} \times (\texttt{thComp} \times \texttt{bbodyrad} + \texttt{diskbb})$. For visual clarity, only data collected by FXT-A in eight out of 22 observations are shown. Moreover, in the upper panel, error bars of the data are omitted and adjacent bins are merged. }
    \label{fig:all_fxt_spec}
\end{figure}
\begin{figure*}
\centering
\includegraphics[width=\textwidth]{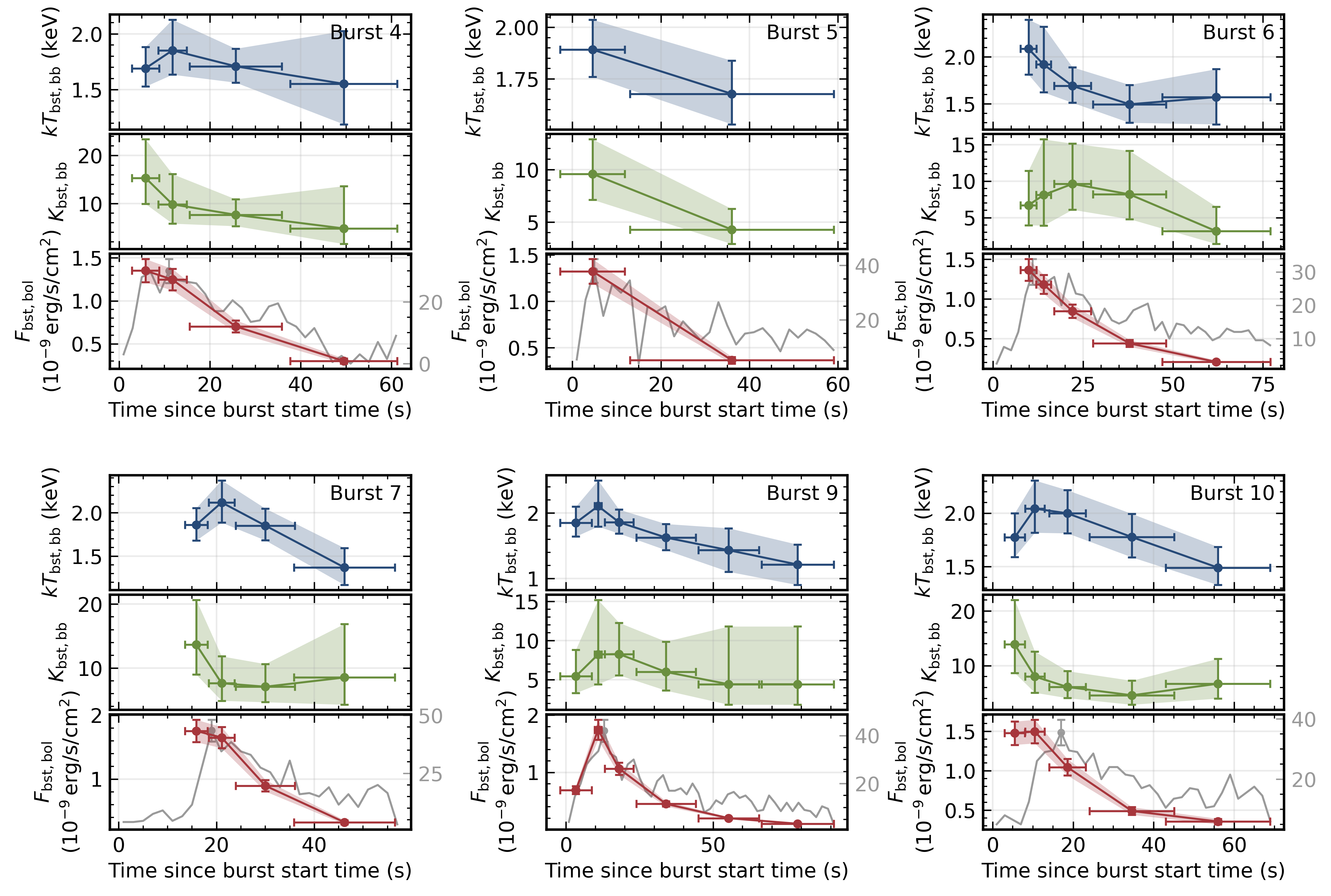}
    \caption{Evolution of key spectral parameters obtained from the time-resolved spectral analysis for six out of seven type-I X-ray bursts detected by \nustar. Results for burst No.~8 are presented in Figure~\ref{fig:burst8}. The \texttt{Xspec} model adopted in the analysis is $\texttt{TBabs} \times (\texttt{nthComp}+\texttt{bbodyrad})$. $T_{\rm bst,bb}$ and $K_{\rm bst,bb}$ are the temperature and normalization of the \texttt{bbodyrad} component, which models the burst emission. $F_{\rm bst,bol}$ is the unabsorbed bolometric flux of the burst emission, which is approximated by the flux of the \texttt{bbodyrad} component, after extrapolating it to the 0.01--100\,keV band. The values of $K_{\rm bst,bb}$ were linked for the last two time segments in burst No.~9 during spectral fitting, because they are otherwise unconstrained. Shaded areas denote uncertainties at 90\% confidence level for $kT_{\rm bst,bb}$ and $K_{\rm bst,bb}$. For $F_{\rm bst,bol}$, we adopted a nominal uncertainty of 10\% of its value. In the bottom panels, the observed \nustar/FPMA light curve (in count rate) of each burst is plotted (in gray) together with $F_{\rm bst,bol}$ for comparison. A representative error bar is given at the peak of each light curve. }
    \label{fig:NuSTAR_burst_time_res}
\end{figure*}
\begin{figure*}
\centering
\includegraphics[width=\textwidth]{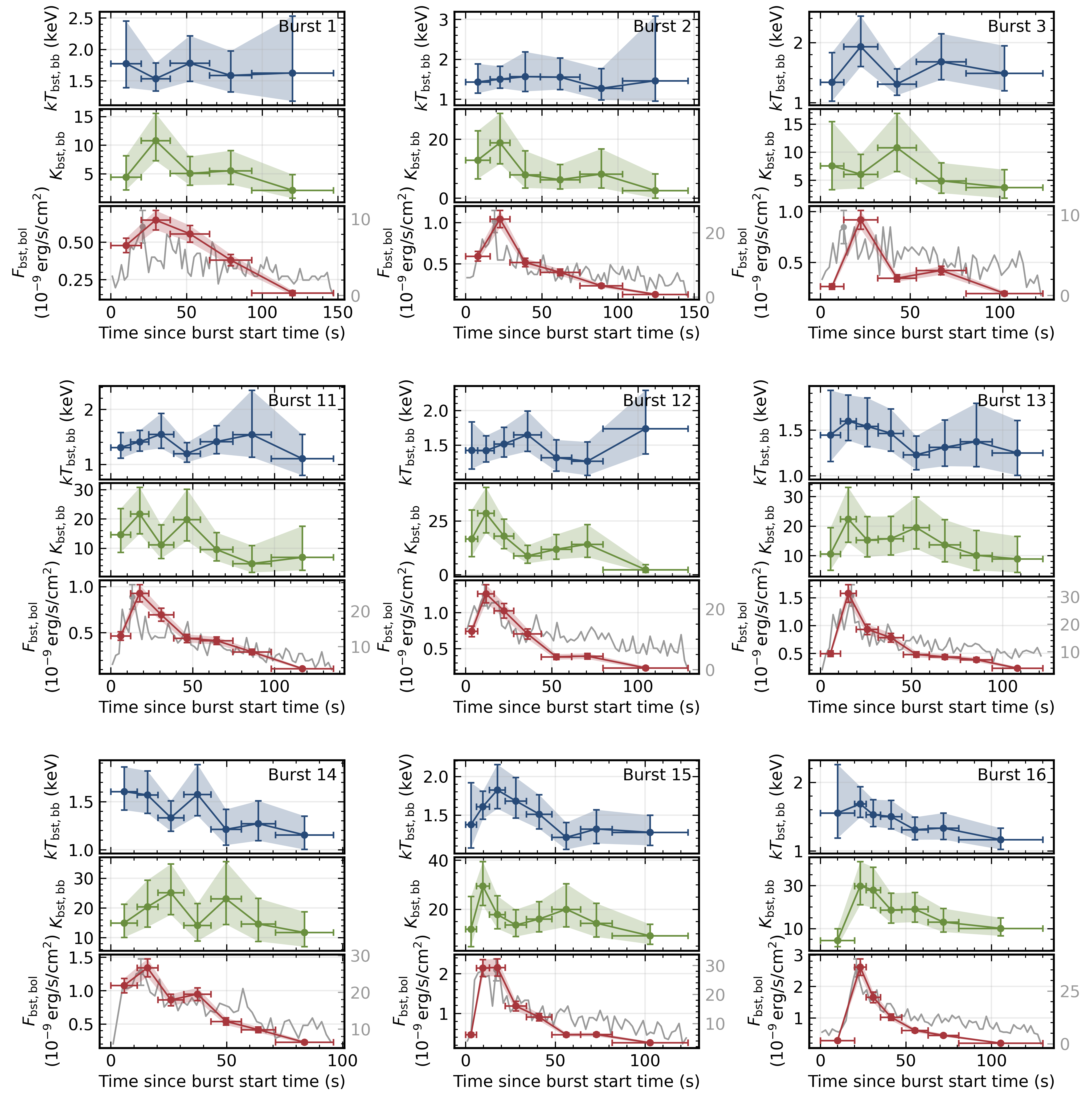}
    \caption{Evolution of key spectral parameters obtained from the time-resolved spectral analysis for nine out of ten type-I X-ray bursts detected by EP/FXT. Results for burst No.~8 are presented in Figure~\ref{fig:burst8}. The \texttt{Xspec} model adopted in the analysis is $\texttt{TBabs} \times \texttt{bbodyrad}$. $T_{\rm bst,bb}$ and $K_{\rm bst,bb}$ are the temperature and normalization of the \texttt{bbodyrad} component, which models the burst emission. $F_{\rm bst,bol}$ is the unabsorbed bolometric flux of the burst emission, which is approximated by the flux of the \texttt{bbodyrad} component, after extrapolating it to the 0.01--100\,keV band. Shaded areas denote uncertainties at 90\% confidence level for $kT_{\rm bst,bb}$ and $K_{\rm bst,bb}$. For $F_{\rm bst,bol}$, we adopted a nominal uncertainty of 10\% of its value. In the bottom panels, the observed FXT-A light curve (in count rate) of each burst is plotted (in gray) together with $F_{\rm bst,bol}$ for comparison. A representative error bar is given at the peak of each light curve. }
    \label{fig:FXT_burst_time_res}
\end{figure*}
\begin{figure*}
\centering
\includegraphics[width=0.9\textwidth]{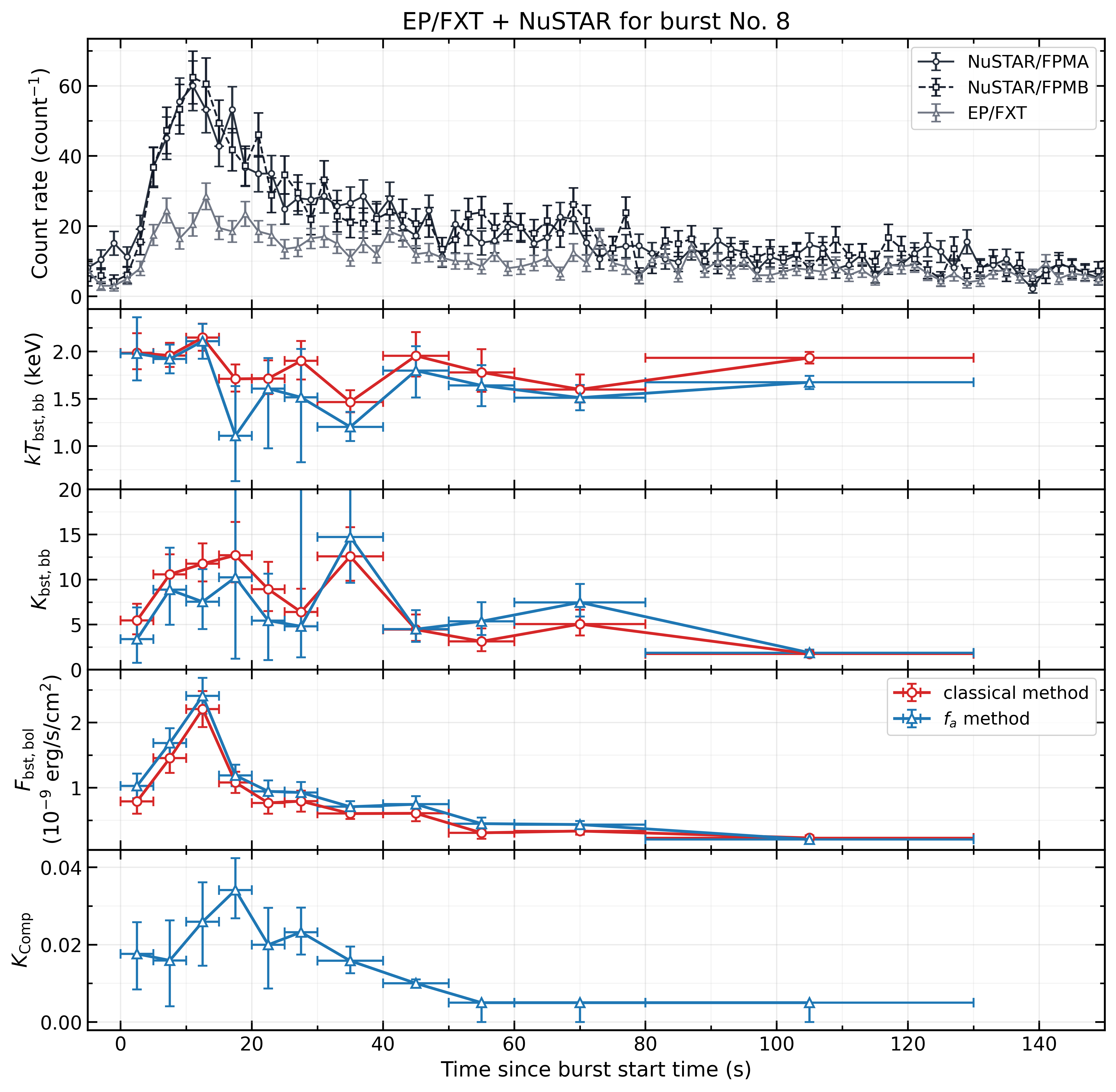}
    \caption{Comparison of the results obtained with the two spectral fitting methods for burst No.~8 observed simultaneously by EP/FXT and \nustar. From top to bottom, the panels show: the 2-s bin light curves from EP/FXT (0.5--10 keV; gray) and \nustar\ (3--50 keV; black); the blackbody temperature ($kT_{\rm bst,bb}$); the blackbody normalization ($K_{\rm bst,bb}$); the bolometric flux ($F_{\rm bst,bol}$; in units of $10^{-9}\,{\rm erg/s/cm^{-2}}$); and the normalization of the \texttt{nthComp} component ($K_{\rm Comp}$). Red circles denote results obtained with the classical method, while blue triangles correspond to the $f_a$ method. }
    \label{fig:burst8}
\end{figure*}

\clearpage

\bibliographystyle{aasjournal}
\bibliography{refs}

\end{document}